\documentclass[fleqn,usenatbib,useAMS]{mnras}

\usepackage[percent]{overpic}
\usepackage{newtxtext,newtxmath}

\usepackage[T1]{fontenc}
\usepackage{ae,aecompl}
\usepackage{multirow}

\usepackage{graphicx, float}	
\usepackage{subcaption} 
\usepackage{parcolumns} 
\usepackage{amsmath}	
\usepackage{url}

\newcommand{\guptainprep}{Gupta et al.\ in prep}

\newcommand{\shannoninprep}{Shannon et al.\ in prep}
\newcommand{\hoffmanninprep}{Hoffmann et al.\ in prep}
\newcommand{\myedit}[1]{\textcolor{black}{#1}}

\title[]{The impact of the FREDDA dedispersion algorithm on $H_0$ estimations with FRBs}   

\author[Hoffmann et al.]{
J.~Hoffmann,${^{1}}$\thanks{E-mail:jordan.hoffmann@postgrad.curtin.edu.au}
C.W.~James,${^{1}}$\thanks{E-mail: clancy.james@curtin.edu.au}
H.~Qiu,${^{2}}$
M.~Glowacki,${^{1}}$
K.W.~Bannister,${^{3}}$ 
V.~Gupta,${^{3}}$ 
J.X.~Prochaska,${^{4,5,6}}$ \newauthor
A.~Bera,${^{1}}$
A.T.~Deller,${^{7}}$ 
K.~Gourdji,${^{7}}$
L.~Marnoch,${^{8,3,9,10}}$ 
S.D.~Ryder,${^{8,9}}$ 
D.R.~Scott,${^{1}}$ 
R.M.~Shannon,${^{7}}$ \newauthor 
N.~Tejos$^{11}$
\\ 
$^{1}$International Centre for Radio Astronomy Research, Curtin University, Bentley, WA 6102, Australia \\
$^{2}$SKA Observatory, Jodrell Bank, Lower Withington,
Macclesfield, SK11 9FT, UK \\
$^{3}$ATNF, CSIRO, Space and Astronomy, PO Box 76, Epping NSW 1710 Australia \\
$^{4}$Department of Astronomy and Astrophysics, University of California, Santa Cruz, CA 95064, USA\\
$^{5}$Kavli Institute for the Physics and Mathematics of the Universe, 5-1-5 Kashiwanoha, Kashiwa 277-8583, Japan\\
$^{6}$Division of Science, National Astronomical Observatory of Japan, 2-21-1 Osawa, Mitaka, Tokyo 181-8588, Japan\\
$^{7}$Centre for Astrophysics and Supercomputing, Swinburne University of Technology, P.O. Box 218, Hawthorn, VIC 3122, Australia \\
$^{8}$School of Mathematical and Physical Sciences, Macquarie University, NSW 2109, Australia \\
$^{9}$Astrophysics and Space Technologies Research Centre, Macquarie University, Sydney, NSW 2109, Australia \\
$^{10}$ARC Centre of Excellence for All-Sky Astrophysics in 3 Dimensions (ASTRO 3D), Australia\\
$^{11}$Instituto de F\'isica, Pontificia Universidad Cat\'olica de Valpara\'iso, Casilla 4059, Valpara\'iso, Chile\\
}

\date{Accepted XXX. Received YYY; in original form ZZZ}

\pubyear{2023}

\begin{document}

\label{firstpage}
\pagerange{\pageref{firstpage}--\pageref{lastpage}}
\maketitle

\begin{abstract}
    Fast radio bursts (FRBs) are transient radio signals of extragalactic origins that are subjected to propagation effects such as dispersion and scattering. It follows then that these signals hold information regarding the medium they have traversed and are hence useful as cosmological probes of the Universe. Recently, FRBs were used to make an independent measure of the Hubble Constant $H_0$, promising to resolve the Hubble tension given a sufficient number of detected FRBs. Such cosmological studies are dependent on FRB population statistics, cosmological parameters and detection biases, and thus it is important to accurately characterise each of these. In this work, we empirically characterise the sensitivity of the Fast Real-time Engine for Dedispersing Amplitudes (FREDDA) which is the current detection system for the Australian Square Kilometer Array Pathfinder (ASKAP). We coherently redisperse high-time resolution data of 13 ASKAP-detected FRBs and inject them into FREDDA to determine the recovered signal-to-noise ratios as a function of dispersion measure (DM). We find that for 11 of the 13 FRBs, these results are consistent with injecting idealised pulses. Approximating this sensitivity function with theoretical predictions results in a systematic error of 0.3$\,$km$\,$s$^{-1}\,$Mpc$^{-1}$ on $H_0$ \myedit{when it is the only free parameter. Allowing additional parameters to vary could increase this systematic by up to $\sim1\,$km$\,$s$^{-1}\,$Mpc$^{-1}$.} We estimate that this systematic will not be relevant until $\sim$400 localised FRBs have been detected, but will likely be significant in resolving the Hubble tension.
\end{abstract}

\begin{keywords}
fast radio bursts -- cosmological parameters
\end{keywords}

\section{Introduction} \label{sec:Introduction}
Fast radio bursts (FRBs) are millisecond-duration, highly energetic signals in the radio spectrum \citep{Lorimer2007, Thornton2013}. Such bursts have been shown to have an extragalactic origin, thus allowing for their use as cosmological probes \citep[e.g.][]{Bannister2019b}. The production mechanism for FRBs is still unknown and progenitor models are bountiful \citep[see][for a review]{platts2019}. Research on host galaxies is expected to give insights into this issue but is presently limited by small sample statistics \citep[e.g.][]{Bhandari2022, Gordon2023}. However, such mysteries do not restrict the use of FRBs in cosmological studies \citep[e.g.][]{Macquart2020, james2022b}.

FRB radiation experiences a frequency-dependent time delay when passing through cold plasmas. Such an effect is quantified by the dispersion measure (DM) which indicates the free electron column density along the line of sight. Correlations between redshift and the DM attributed to cosmological sources such as the intergalactic medium (IGM) and intervening structures then inform us about the cosmology of our Universe. It follows then that FRBs localised to host galaxies (which have a corresponding redshift) hold one of the most constraining powers in cosmological studies \citep[e.g.][]{Macquart2020, james2022b}. This makes raadio-telescopes with arcsecond localisation capabilities, such as the Australian Square Kilometer Array Pathfinder \citep[ASKAP;][]{Hotan2021}, the Deep Synoptic Array \citep[DSA;][]{dsa102019} and MeerKAT \citep{jonas2016}, important instruments for these studies. 

\citet{Macquart2020} used five ASKAP localised FRBs to derive a value for the cosmic baryon density and solve the `missing baryons problem' \citep{fukugita1998}, demonstrating the power of FRBs in solving outstanding cosmological mysteries. The next cosmological issue that FRBs may be able to address is the Hubble tension. Early time and late time measurements of the Hubble constant $H_0$ have shown significant discrepancies: observations of the Cosmic Microwave Background (CMB) by \textit{Planck} gave a value of $67.4 \pm 0.5$\,km\,s$^{-1}$\,Mpc$^{-1}$ \citep{Planck2018} while measurements using local distance ladders gave a value of $73.04 \pm 1.04$$\,$km$\,$s$^{-1}\,$Mpc$^{-1}$ \citep{SH0ES2021}. \citet{james2022b} recently gave an estimate of $73^{+12}_{-8}$ km s$^{-1}$ Mpc$^{-1}$ for $H_0$ using 16 ASKAP-localised FRBs, demonstrating the possibility of relieving the Hubble tension. Small-sample statistics and poor constraining power on other model parameters \citep[e.g.][]{Baptista2023} currently limit the precision of such an estimation, however, as the statistical error in cosmological parameter estimation decreases with new FRB detections, the relative importance of systematic errors will increase. One such systematic error is the instrumental detection bias against high-DM FRBs due to the smearing of the pulse. Here, we aim to characterise this bias for FRBs detected under the Commensal Real-time ASKAP Fast Transients (CRAFT) survey. 

The search pipeline used by the CRAFT collaboration is the Fast Real-time Engine for Dedispersing Amplitudes \citep[FREDDA;][]{Bannister2019}. FREDDA is a GPU implementation of the Fast Dispersion Measure Transform \citep[FDMT;][]{Zackay2017} which allows it to quickly search the data over a large range of trial DMs in real-time. \citet{Qiu2023} recently profiled FREDDA's sensitivity as a function of DM by injecting pulses that were Gaussian in time with a flat, broadband spectrum embedded in Gaussian noise. While such an analysis is a clear improvement on theoretical models, it does not consider how the intricate and varied morphology of FRBs, RFI or spectral dependence of background noise affect the recovered sensitivity.

In this work, we focus on characterising the sensitivity of FREDDA with the inclusion of real burst morphologies. The formalism for the sensitivity is presented in Section \ref{sec:sensitivity_function_formalism}. For the first time, we coherently re-disperse actual CRAFT-detected FRBs from high-time resolution (HTR) voltages \citep{Cho2020, CELEBI}. This produces the burst that would have been detected had it passed through a different electron column density. We then reinject these bursts into FREDDA at varying DMs to empirically determine the signal-to-noise ratio (SNR) as a function of DM. The full method for this analysis is given in Section \ref{sec:FREDDA_injection}. Resulting sensitivity functions are also presented alongside relevant discussion. In Section \ref{sec:impacts_on_H0_calculations} we discuss the systematic errors introduced in $H_0$ estimations due to idealised $\eta(\mathrm{DM})$ values. Lastly, we conclude our findings in Section \ref{sec:conclusions}.

\section{Sensitivity function formalism} \label{sec:sensitivity_function_formalism}
The fluence of a burst is defined as the integral of the flux across the duration of the burst and characterises the total amount of energy density contained in it. For a given fluence, temporally wider bursts integrate over a larger quantity of noise and therefore have a lower signal-to-noise ratio (SNR), resulting in a higher fluence threshold. This decrease in sensitivity is accounted for by specifying an efficiency function $\eta$. We normalise $\eta$ such that it represents the relative sensitivity to an idealised 1 ms wide burst. That is,
\begin{eqnarray}
     \eta = \frac{\mathrm{SNR_{eff}}}{\mathrm{SNR_{1ms}}}\,, \label{eq:SNR-eta}
\end{eqnarray}
where SNR$_{\mathrm{eff}}$ is the effective SNR and SNR$_{\mathrm{1ms}}$ is the SNR of a 1 ms burst \citep{Cordes2003}. FRB population models typically determine $\eta$ as a function of the effective width $w_{\mathrm{eff}}$ of the FRB \citep[e.g.][]{Gardenier2019}. For an ideal case, the SNR for a burst of constant fluence will scale inversely proportionally to the square root of $w_{\mathrm{eff}}$. Thus, the sensitivity function can be approximated by
\begin{eqnarray}
    \eta = \sqrt{\frac{1 \: \mathrm{ms}}{w_{\mathrm{eff}}}}\,,
\end{eqnarray}
where 1 ms is an effective width. Algorithmic effects will cause deviations from this due to assumptions of the search filter shape and ambiguities in the position of each FRB within each time bin. To determine $w_{\mathrm{eff}}$, one must account for the intrinsic emission width $w_{\mathrm{int}}$, the scattering timescale $\tau$, the temporal resolution used in the search algorithm $w_{\mathrm{res}}$ and DM smearing $w_{\mathrm{smear}}$ \citep{Cordes2003, Arcus2020}. Current studies indicate that scattering and DM are not correlated \citep{Chawla2022, Gupta2022} and hence $\tau$ and $w_{\mathrm{int}}$ are usually combined into some incident width $w_{\mathrm{inc}}$. DM smearing refers to dispersion between the top and bottom of a given frequency channel, therefore causing a broadening of the FRB. For a given spectral channel, the amount of intrachannel smearing is approximated by
\begin{eqnarray}
    t_{\mathrm{smear}} = \frac{2 \mathcal{D} \Delta \nu}{\nu_c^3} \mathrm{DM}\,, \label{eq:smear}
\end{eqnarray}
where $\mathcal{D} \cong e^2 / (2 \pi m_{\mathrm{e}} c) \approx 4.148808 \times 10^3$ MHz$^2$\,pc$^{-1}$\,cm$^3$\,s \citep{pulsarhandbook}, $\Delta \nu$ is the channel width and $\nu_c$ is the central frequency of the channel. The average smearing over the band $w_{\mathrm{smear}}$ is then approximated by the smearing in the central frequency bin. This smearing term gives a DM dependence to $\eta$ and hence can impact cosmological models which fundamentally aim to calculate $p(z,\mathrm{DM})$ -- The probability of detecting an FRB at a given redshift and DM. Thus, it is important to account for such an effect to minimise systematic errors.

Theoretical predictions of \citet{Cordes2003} suggest that $w_{\mathrm{eff}}$ is a quadratic summation of each contributing factor and is therefore given by
\begin{eqnarray}
    w_{\mathrm{eff}} = \sqrt{w_{\mathrm{smear}}^2 + w_{\mathrm{res}}^2 + w_{\mathrm{inc}}^2}\,.
\end{eqnarray}
\citet{Arcus2020} instead takes a linear combination giving
\begin{eqnarray}
    w_{\mathrm{eff}} = c_1 w_{\mathrm{smear}} + c_2 w_{\mathrm{res}} + w_{\mathrm{inc}}\,,
\end{eqnarray}
where $c_1$ and $c_2$ are fitting parameters specific to each telescope. The requirement to fit $c_1$ and $c_2$ for each telescope limits the genericity of the \citet{Arcus2020} model and hence the default implementation of \citet{james2022b} uses the \citet{Cordes2003} model. Regardless, both of these models assume an ideal scenario, ignoring algorithmic effects and any temporal and/or spectral structures of the FRB. As such, many collaborations have characterised the sensitivity of their detection system using pulse injection techniques \citep[e.g.][]{CHIME2018, Agarwal2020, Gupta2021, Qiu2023} in which mock FRBs are injected into the detection system and the recovered sensitivity is directly estimated. 

\section{FREDDA injection} \label{sec:FREDDA_injection}
In this section, we describe the method by which we obtained the sensitivity as a function of DM specific to a given FRB.

\subsection{Redispersing HTR voltages} \label{sec:redispersing_HTR_voltages}
For a given FRB detected by ASKAP, we use the high-time resolution (HTR) data produced via the CELEBI pipeline \citep{CELEBI}. The sample of FRBs that we discuss in this paper corresponds to all FRBs processed with CELEBI at the time of publication. We begin with the time series of the coherently dedispersed complex voltages for each of the two antenna polarisations. These time series have a duration of 3.1 seconds and a temporal resolution of (336 MHz)$^{-1}\approx$ 3 ns. The data is coherently beamformed and hence does not perfectly replicate the data on which FREDDA typically operates. While it is possible to directly use the antenna voltages, it is far less convenient and we expect the analysis to be equivalent. The largest differences will be an increase in the absolute SNR proportional to $\sqrt{N_{\rm antennas}}$ (assuming perfect coherence) and a change in the RFI environment of the data. We ultimately scale the SNR to a value of $\eta$ and find that varying RFI environments do not have a large impact on the recovered sensitivities (see Section \ref{sec:sensitivity_results}) and hence these effects were ignored. We extract $\sim$1 second of data around the FRB pulse region to minimise computational memory usage. The corresponding frequency spectrum for each antenna polarisation is then constructed by performing a complex-to-complex fast Fourier transform (FFT). 

A dispersion of the desired magnitude is coherently applied to the voltages in the frequency domain relative to the highest observational frequency. Dispersion relative to the highest frequency of the band can be described by a transfer function in the frequency domain such that
\begin{eqnarray}
    V_{\mathrm{dispersed}}(\nu) = V(\nu)\mathrm{exp} \left(i \: \frac{2 \pi \mathcal{D} \mathrm{DM} (\nu_{\mathrm{max}} - \nu)^2}{\nu_{\mathrm{max}}^2 \nu}\right)\,,
\end{eqnarray}
where $V(\nu)$ and $V_{\mathrm{dispersed}}(\nu)$ represent the complex voltages in the frequency domain before and after dispersion and $\nu_{\mathrm{max}}$ is the highest frequency of the band \citep{pulsarhandbook}. The coherent redispersion is applied by multiplying the frequency series of both antenna polarisations by this transfer function. We then perform a 336-point complex-to-complex FFT to recover the 336 1 MHz coarse spectral channels. We determine the overall intensity $I$ via
\begin{eqnarray}
    I = |V_x(t,\nu)|^2 + |V_y(t,\nu)|^2\,,
\end{eqnarray}
where $V_x$ and $V_y$ represent the redispersed voltages for each of the polarisations. We then integrate this dynamic spectrum in time to reproduce the coarse temporal resolution of data on which FREDDA operated at the time of detection, typically of order $\sim$1 ms. ASKAP implements a polyphase filterbank (PFB) rather than an FFT to minimise spectral leakage. These PFBs produce 784 overlapping coarse channels with the central frequency of each channel separated by 1 MHz and the width of each channel being 32/27 MHz \citep{Hotan2021}. Of these, only 336 channels are used in FRB searches and consequently saved in the voltage buffers. Hence, it is not possible to perfectly reconstruct the original voltage data. To account for this, we implemented a basic PFB algorithm. The implemented PFB used a sinc window with a period corresponding to the bandwidth (336 samples) of observations multiplied by a sine envelope to taper the edges. The window spans 8$\times 336 = 2688$ samples. The inclusion of this PFB increased computational costs significantly but resulted in no significant change and therefore was not utilised in the results we present.

The dynamic spectrum was then rescaled to have a mean of 128 and a standard deviation of 8 and was converted to an 8-bit integer format. Some FRBs have single channels containing strong RFI that completely dominate all others. As such, normalisation with the inclusion of these channels results in the remaining spectra being reduced to 0 in the 8-bit integer format. For these FRBs, it was thus necessary to mask these channels prior to rescaling.

The dynamic spectrum is padded with 5 seconds of frequency-dependent noise, allowing FREDDA to obtain a baseline noise level. This noise is assumed to be Gaussian and is randomly generated using the mean and standard deviation of each frequency channel in an off-pulse region. The introduction of randomly generated noise and the use of relatively coarse DM trials induces variations in the recovered SNR of up to $\pm 2$. The padded spectrum is then saved to a filterbank file and injected into an offline version of FREDDA to identify any pulse candidates with their associated SNR. This process was repeated for a number of trial DMs ranging from 0 pc cm$^{-3}$ to the DM corresponding to a time delay across the entire band of 4096 time bins (the assigned search limit for FREDDA due to computational restrictions) in steps of 50 pc cm$^{-3}$.

\begin{figure*}
\begin{center}
\begin{subfigure}{8.7cm}
  \centering
  \includegraphics[width=8.7cm]{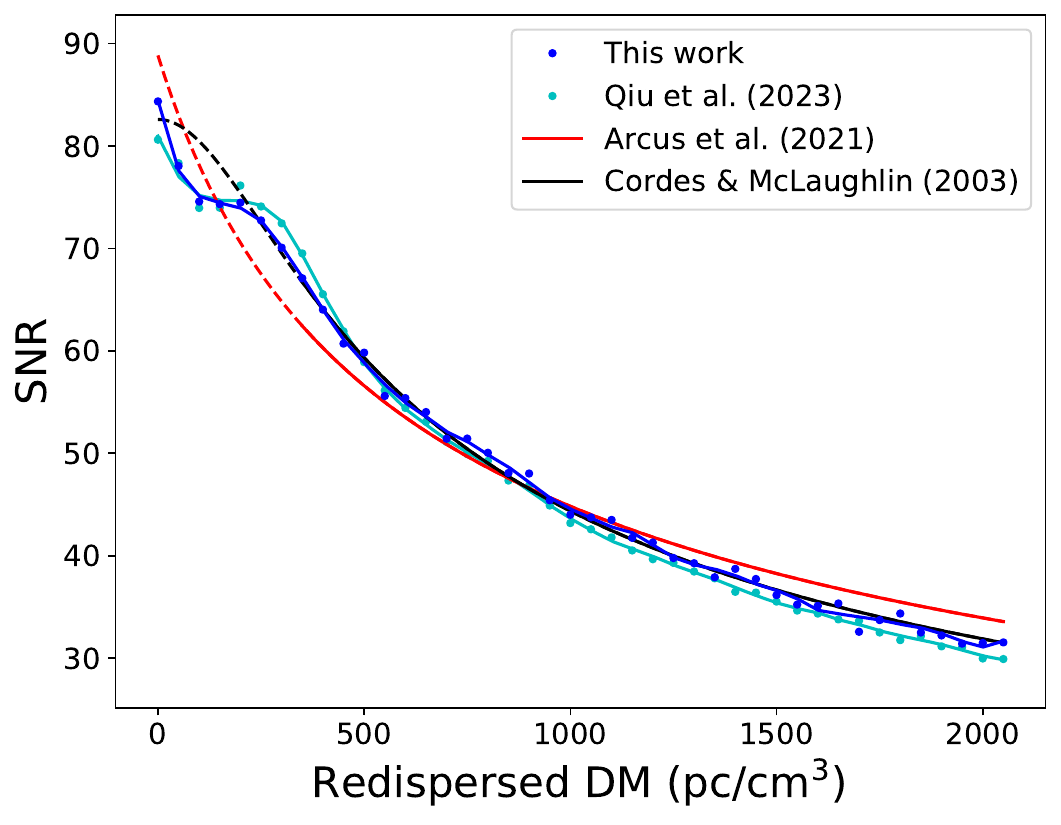}
\end{subfigure}%
\begin{subfigure}{7.3cm}
  \centering
  \includegraphics[width=7.3cm]{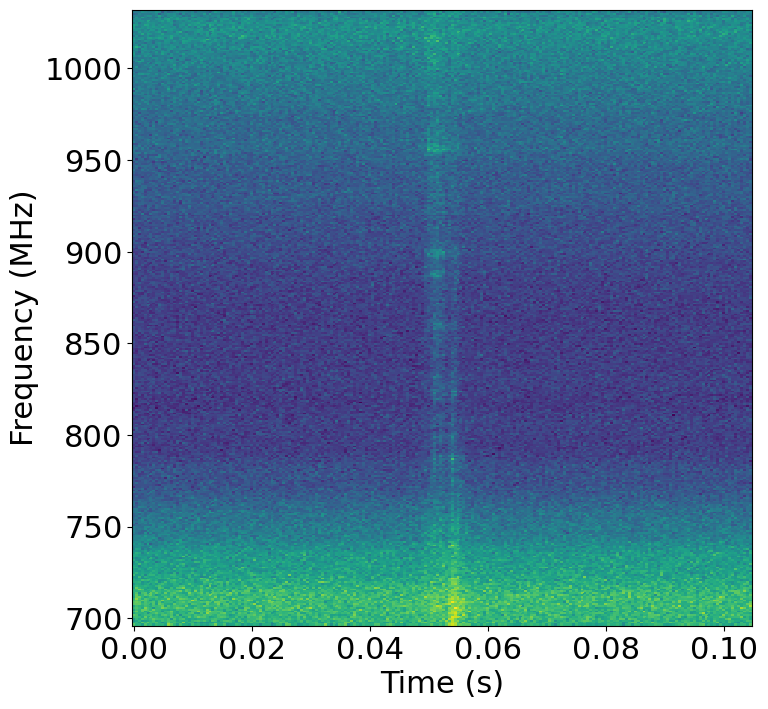}
\end{subfigure}
\caption[]{Left: The SNR recovered from FREDDA for FRB 20220501C (dark blue points) smoothed with a Savitzky-Golay filter (dark blue line). Theoretical predictions from \citet{Cordes2003} (black line) and \citet{Arcus2020} (red line) are overplotted. The solid regions of the lines show the DM range over which fitting between the theoretical models and data points was completed. Pulse injection results (light blue points) from \citet{Qiu2023} for Gaussian pulses of a similar width to the FRB and in a similar frequency band are also shown. Right: The FRB's dynamic spectrum without RFI flagging or channel-by-channel normalisation.}
\label{fig:220501_SNR}
\end{center}
\vspace{-3ex}
\end{figure*}

\subsection{Model predictions and plotting} \label{sec:models}
Relevant properties of all FRBs to which this analysis was applied are given in Table \ref{tab:FRB_properties}. Figure \ref{fig:220501_SNR} shows our recovered sensitivity function for FRB 20220501C alongside its dynamic spectrum as an example. The dark blue points show the SNRs produced by FREDDA. The light blue points were produced by rerunning the pulse injection of \citet{Qiu2023} with the same resolution and frequency band as the detected FRB. These points were scaled to approximately match the scale of the dark blue points. The full-width half-maximum (FWHM) of the pulse was taken to be the same as $w_{\mathrm{inc}}$. The dark and light blue lines are numerically smoothed versions of the corresponding data points using a Savitzky-Golay filter \citep{savgol}. We exclude the 0 pc cm$^{-3}$ point for FRBs with $w_{\mathrm{inc}} \ll w_{\mathrm{res}}$ as at 0 pc cm$^{-3}$, the SNR does not depend on the start time of the burst, but at all other DMs, the phase of the start time with respect to the sampling time dictates the best-fitting template in time--frequency space. This causes a discontinuity in the SNR by a factor of $\sim$$\sqrt{2}$ between the 0 pc cm$^{-3}$ and 50 pc cm$^{-3}$ (our first DM trial) points (\guptainprep). The black and red lines show the \citet{Cordes2003} and \citet{Arcus2020} models respectively. These models require an estimation of $\mathrm{w_{inc}}$. However, predicting such a width is difficult due to complex pulse morphologies and detection system peculiarities. The reported FWHM of the pulse did not produce results consistent with the theoretical expectation. Hence, we approximate $w_{\mathrm{inc}}$ by a maximum likelihood fit of the \citet{Cordes2003} model for $w_{\mathrm{inc}}$ at DM values larger than some visually determined cutoff. The solid portions of the model curves represent this region. Often this width was sensitive to the cutoff DM and hence was used as a base estimate but was adjusted visually. The values of $\mathrm{w_{inc}}$ used are given in Table \ref{tab:FRB_properties}. We also fit the scaling factor (SNR$_{1\mathrm{ms}})^{-1}$ from Equation \eqref{eq:SNR-eta} simultaneously which gives the ratio of $\eta$(DM) to SNR$_{\mathrm{eff}}$. Therefore, the \citet{Cordes2003} and \citet{Arcus2020} models shown here are better fits than they would be if coherently redispersed HTR data was absent, as is the case for all existing applications in the literature.

\subsection{Old FREDDA versions} \label{sec:old_fredda}
The first FRB detected by FREDDA was in 2017 \citep{Shannon2018} and since then FREDDA has been under development coincident with its operational use. When determining the response of FREDDA for use in cosmological studies, it is necessary to use the response of the detection algorithm which was operational at the time. Most developments did not cause significant variations in the response function and hence do not need to be considered, however, there are broadly three versions in which significant differences are present.

`Version 1' was used prior to August $30^{\rm th}$ 2019. It did not calculate the effect of dispersion-measure smearing within coarse channels, and summed over only a single time bin for each DM trial. Such smearing was effectively taken into account using a width search, such that the search templates were effectively boxcars of equal length applied equally to each coarse channel. `Version 2' was used until April $6^{\rm th}$ 2020, and implemented channel-specific time-domain sums to account for DM smearing. However, this meant that DM trials would be correlated for sequential time samples, since the DM smearing times overlapped. However, the width search did not account for this, meaning that incorrectly high SNR values were returned. This was fixed in `Version 3' --- the version described by \citet{Qiu2023}, and the default when referring to `FREDDA' --- which accounted for this correlation in calculating SNR values. This later version has been running since April $6^{\rm th}$ 2020.

\begin{figure}
\begin{center}
\includegraphics[width=\columnwidth]{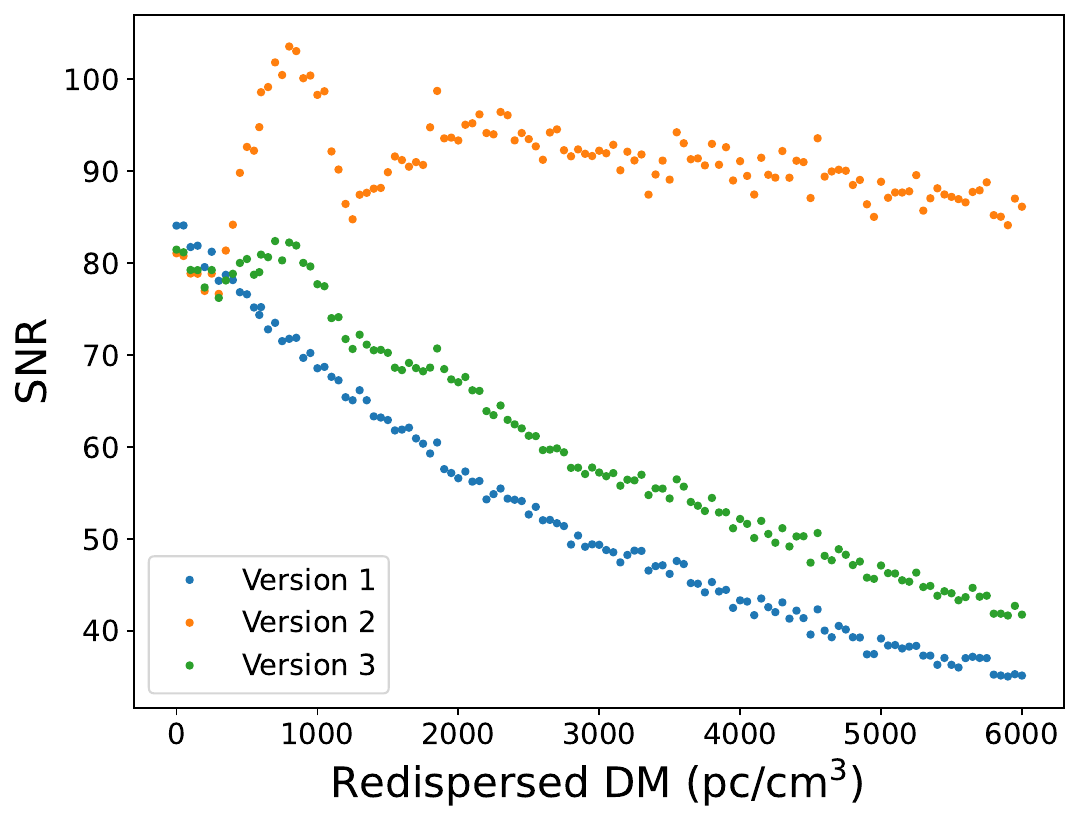}
\caption[]{Recovered SNRs from injecting redispersed data of FRB 20190711A using versions of FREDDA from three different time periods. The blue dots show the original response when DM smearing was not considered. The orange dots show the response when DM smearing was considered but the SNR was not correctly calculated. The green points show the response from the current detection system where DM smearing is considered and the SNR is correctly calculated. Further explanations of each version are given in Section \ref{sec:old_fredda}.}
\label{fig:190711_reverted}
\end{center}
\vspace{-3ex}
\end{figure}

\begin{figure*}
\begin{center}
\begin{subfigure}{8.7cm}
  \centering
  \includegraphics[width=8.7cm]{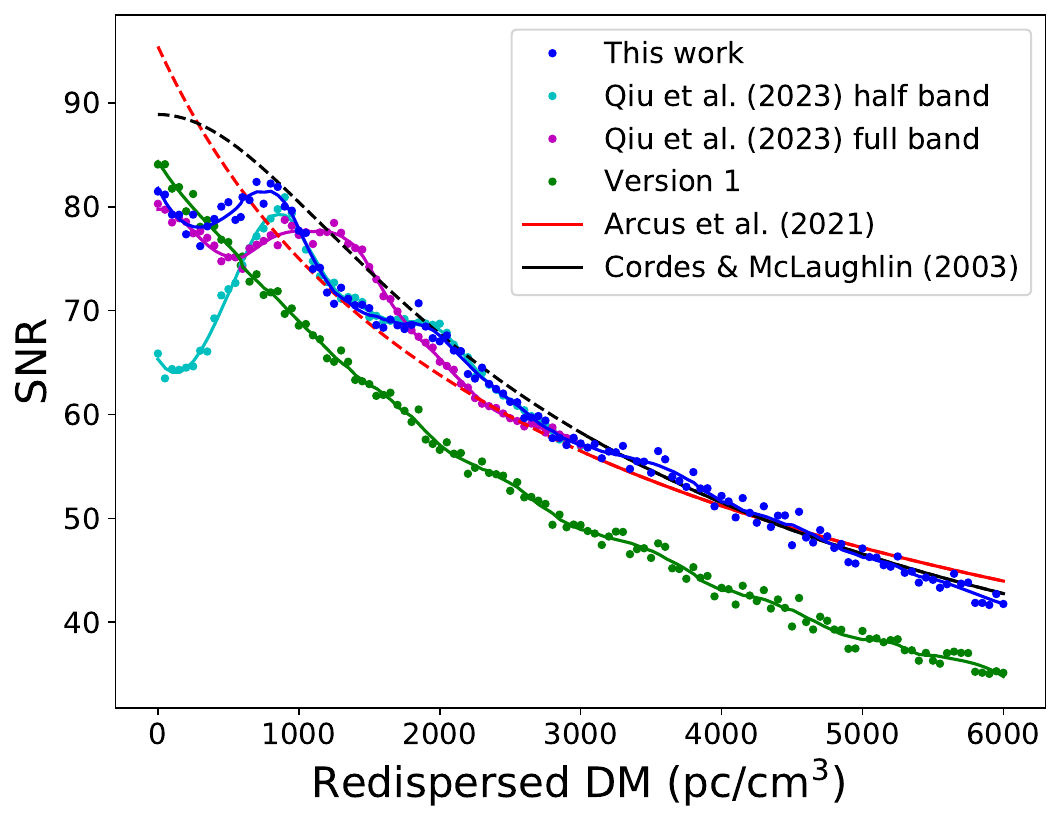}
\end{subfigure}%
\begin{subfigure}{7.3cm}
  \centering
  \includegraphics[width=7.3cm]{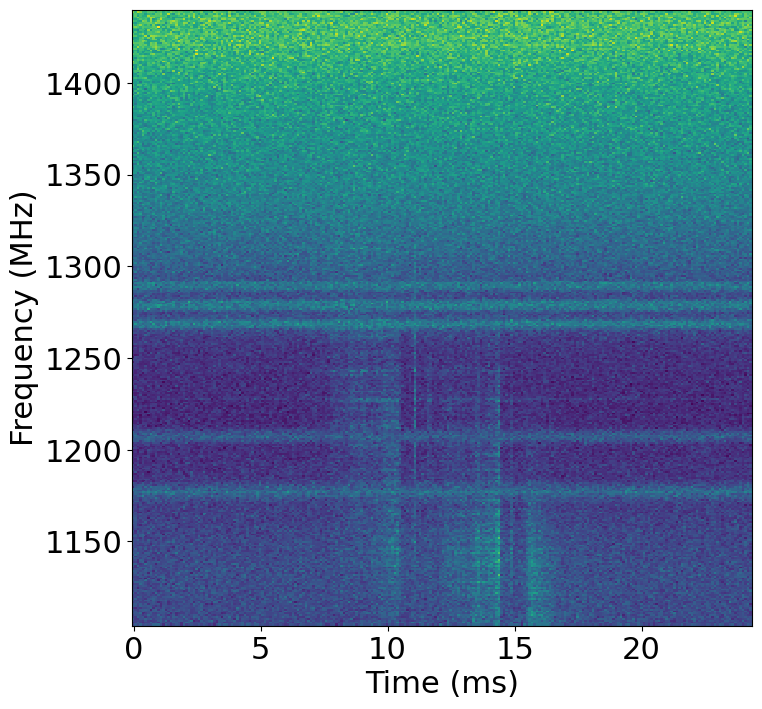}
\end{subfigure}
\caption[]{Same as Figure \ref{fig:220501_SNR} for FRB 20190711A. Two results from pulse injection are also shown. The magenta curve shows the response of a pulse with full band occupancy and the light blue shows an identical pulse only filling the bottom half of the band. The half-band result shows a more accurate representation of this FRB.}
\label{fig:190711_SNR}
\end{center}
\vspace{-3ex}
\end{figure*}

Figure \ref{fig:190711_reverted} shows the response of FRB 20190711A with the three versions of FREDDA as an example. `Version 1' has a reduced (but correctly calculated) SNR at high DM values and `Version 2' overestimates the SNR values at large DMs, which is consistent with expectation. In Figures \ref{fig:190711_SNR} and \ref{fig:SNRs}, FRBs detected before April $6^{\rm th}$ 2020 have the response of the version running at detection shown in green.

\subsection{Comparing sensitivity functions} \label{sec:sensitivity_results}
Sensitivity functions for each of the FRBs are given in Figure~\ref{fig:SNRs} where otherwise not presented in the main text. These figures show the same results as Figure~\ref{fig:220501_SNR} for the other FRBs in question and hence the same analysis discussed previously was used. 

The pulse injection of \citet{Qiu2023} superimposes spectrally uniform pulses which are Gaussian in time onto white noise. Conversely, the FRBs in question contained a variety of widths, scintillation, scattering, intrinsic spectral structure, narrow-band RFI, and spectrally-dependent noise. Some temporal structure exists; however, for most instances, it is negligible at the integrated resolution of $w_{\mathrm{res}}$. Despite the variety of morphologies and noise structures, for an appropriate $w_{\mathrm{inc}}$, good agreement with the scaled pulse injection of \citet{Qiu2023} was observed for the majority of FRBs. We observe a sharp decrease in the SNR values from a DM of 0 pc cm$^{-3}$ to 50 pc cm$^{-3}$ in FRBs 20181112A, 20210320A and 20210407E as predicted for narrow bursts (\guptainprep; see \ref{sec:models} for a brief explanation). An oscillatory behaviour is also present that matches the expectations of \citet{Qiu2023} and is attributed to the FDMT algorithm producing a search template that imprecisely reproduces the DM sweep in a DM-dependent manner. Such an effect is suppressed for larger DMs as less power is contained in each bin (due to DM smearing) and hence the inclusion or exclusion of a bin holds less significance. Neither of these effects are considered in the theoretical models and hence cause deviations from the predicted sensitivity which are more apparent at lower DMs (typically <1000 pc$\,$cm$^{-3}$ but variable depending on the frequency band and integration time). Differences between the \citet{Qiu2023} method and our results are evident in FRBs which have scattering tails that are significant even at the integrated time resolution. Such a discrepancy is expected as the burst structures deviate from the idealised Gaussian pulse.

FRB 20190711A shows the most temporal structure even at the integrated resolution of $w_{\mathrm{res}}$ and exhibits half-band occupancy. This FRB shows the greatest deviations from pulse injection predictions. Figure \ref{fig:190711_SNR} shows the sensitivity functions for this FRB alongside its dynamic spectrum. The magenta curve shows the \citet{Qiu2023} pulse injection in which significant discrepancies in the shape of the sensitivity function are evident. As this FRB only occupies the lower half of the band, we also consider pulse injection of an idealised pulse which similarly fills only half the band with results shown in light blue. The shape of the recovered sensitivity function is more consistent with our results, however, the sensitivity at low DMs is significantly reduced. Due to the complex temporal structure of the FRB, it is unsurprising that the sensitivities are not perfectly described by either of these models.

\begin{figure*}
\begin{center}
\begin{subfigure}{8.7cm}
  \centering
  \includegraphics[width=8.7cm]{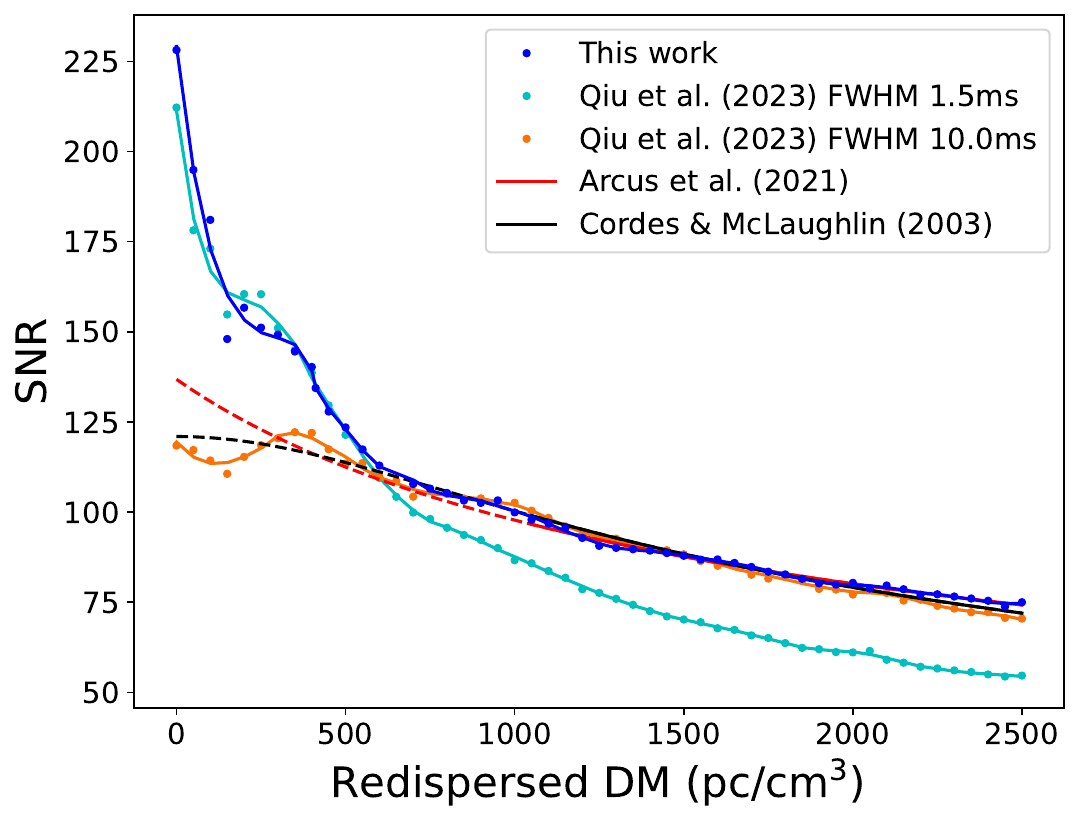}
\end{subfigure}%
\begin{subfigure}{7.3cm}
  \centering
  \includegraphics[width=7.3cm]{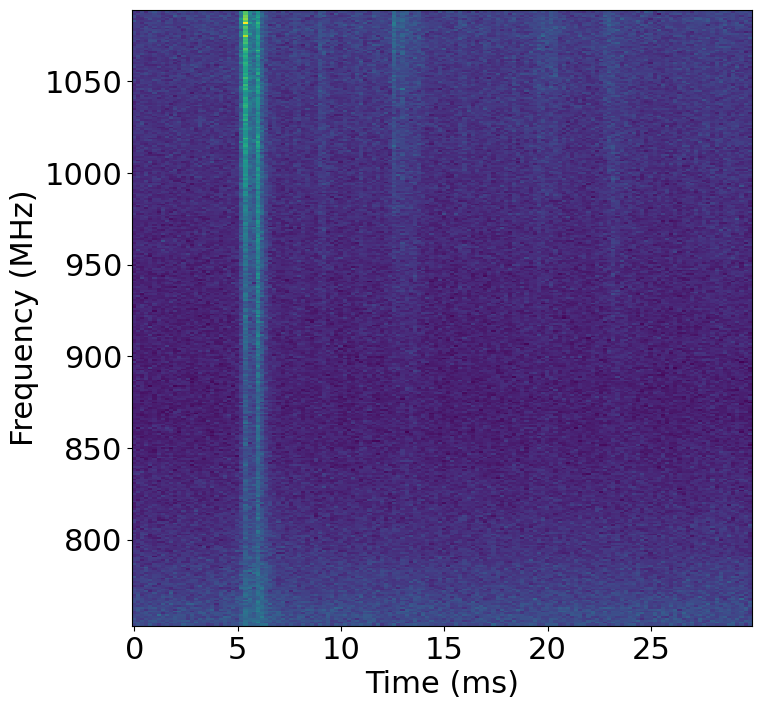}
\end{subfigure}
\caption[]{Same as Figure \ref{fig:220501_SNR} for FRB 20230708A. The lighter blue and orange curves show pulse injection results for Gaussians with the indicated FWHM. The models of \citet{Cordes2003} and \citet{Arcus2020} assume a 10 ms incident width.}
\label{fig:230708_SNR}
\end{center}
\vspace{-3ex}
\end{figure*}

Figure \ref{fig:230708_SNR} shows the obtained sensitivity functions for FRB 20230708A. This FRB has multiple peaks spread over a $\sim$20 ms time interval. For low DM values, FREDDA only identifies the primary peak and hence detects the FRB as a narrow pulse with $w_{\mathrm{inc}} \approx 1$ ms. As such, good agreement with pulse injection of a 1.5 ms FWHM pulse is observed. As DM smearing becomes more significant at larger DMs, FREDDA instead prefers $w_{\mathrm{inc}} \approx 10$ ms and hence shows agreement with pulse injection of a 10 ms FWHM pulse.

In general, the sensitivity of FREDDA for a majority of the analysed FRBs is consistent with that of the idealised pulses of \citet{Qiu2023} given an appropriate width. Of the 13 analysed FRBs, three were detected with `Version 1' of FREDDA and of these only FRB 20190711A shows significant differences compared to the current version of FREDDA due to its large incident width. FRB 20191228A was detected with `Version 2' which shows significant differences in the sensitivity function. The remaining nine were all detected with the most recent version of FREDDA. The consistency we find with results from pulses injected into white noise with no RFI present suggests that the RFI environment at detection has minimal impact on the shape of the sensitivity curve. This agreement also suggests that frequency structures show minimal impact apart from a case in which there is partial band occupancy in which noticeable deviations are present. Fine temporal structures are generally unresolved at the integrated time resolutions and also seem to have a minimal impact on the sensitivity function. 

\begin{table*}
\begin{center}
\caption{Detection properties of FRBs considered in this work. Given are the FRB name, structure maximising DM, estimated DM$_{\mathrm{ISM}}$ from the NE2001 model \citep{Cordes2002}, central observational frequency $\nu_{\mathrm{c}}$, SNR at detection, the optimised incident width $w_{\mathrm{inc}}$, the temporal resolution used in the search $w_{\mathrm{res}}$, redshift $z$, the version of FREDDA operational at the time of detection and reference paper.}
\label{tab:FRB_properties}
\begin{tabular}{llllllllcl}
\hline
TNS name \hspace{1ex} & DM \hspace{1ex} & DM$_{\mathrm{ISM}}$ \hspace{1ex} & $\nu_{\mathrm{c}}$ \hspace{1ex} & SNR \hspace{1ex} & $w_{\mathrm{inc}}$ \hspace{1ex} & $w_{\mathrm{res}}$ \hspace{1ex} & $z$ \hspace{1ex} & Version \hspace{1ex} & Reference \\
& (pc cm$^{-3}$) \hspace{1ex} & (pc cm$^{-3}$) \hspace{1ex} & (MHz) \hspace{1ex} & & (ms) \hspace{1ex} & (ms) \hspace{1ex} & & & \\
\hline 
\vspace{-2ex} \\
20181112A & 589.265 & 40.2 & 1297.5 & 19.3 & 0.03 & 0.864 & 0.4755 & 1 &\citet{Prochaska2019b}\\
\hline
20190611B & 322.22 & 57.6 & 1271.5 & 9.3 & 1.5 & 1.728 & 0.378 & 1 & \multirow{2}{*}{\citet{Macquart2020}} \\
20190711A & 587.77 & 56.6 & 1271.5 & 23.8 & 5.5 & 1.728 & 0.522 & 1 &\\
\hline
20191228A & 296.948 & 32.9 & 1271.5 & 22.9 & 7.8 & 1.728 & 0.243 & 2 & \citet{Bhandari2022} \\
\hline
20200430A & 379.759 & 27.0 & 864.5 & 15.7 & 9.7 & 1.728 & 0.161 & 3 & \citet{Heintz2020} \\
\hline
20210117A & 729.1 & 34.4 & 1271.5 & 27.1 & 1.7 & 1.182 & 0.214 & 3 & \citet{Bhandari2023} \\
\hline
20210320A & 384.59 & 42.2 & 864.5 & 15.3 & 0.21 & 1.728 & 0.28 & 3 & \multirow{2}{*}{\shannoninprep} \\
20210407E & 1784.86 & 154.0 & 1271.5 & 19.1 & 0.65 & 1.182 & -- & 3 & \\
\hline
20210912A & 1233.69 & 30.9 & 1271.5 & 31.7 & 0.05 & 1.182 & -- & 3 & \citet{Marnoch2023} \\
\hline
20220501C & 449.26 & 30.6 & 863.5 & 16.1 & 3.7 & 1.182 & 0.381 & 3 & \multirow{4}{*}{\shannoninprep} \\
20220725A & 288.37 & 30.7 & 920.5 & 12.7 & 1.8 & 1.182 & 0.1926 & 3 & \\
20230526A & 316.148 & 50.0 & 1271.5 & 22.1 & 2.0 & 1.182 & 0.157 & 3 & \\
20230708A & 411.51 & 50.2 & 920.5 & 31.5 & 1.5/10.0 & 1.182 & 0.105 & 3 & \\
\hline
\end{tabular}
\end{center}
\end{table*}

\section{Impacts on $H_0$ calculations} \label{sec:impacts_on_H0_calculations}
\begin{figure}
\begin{center}
\includegraphics[width=\columnwidth]{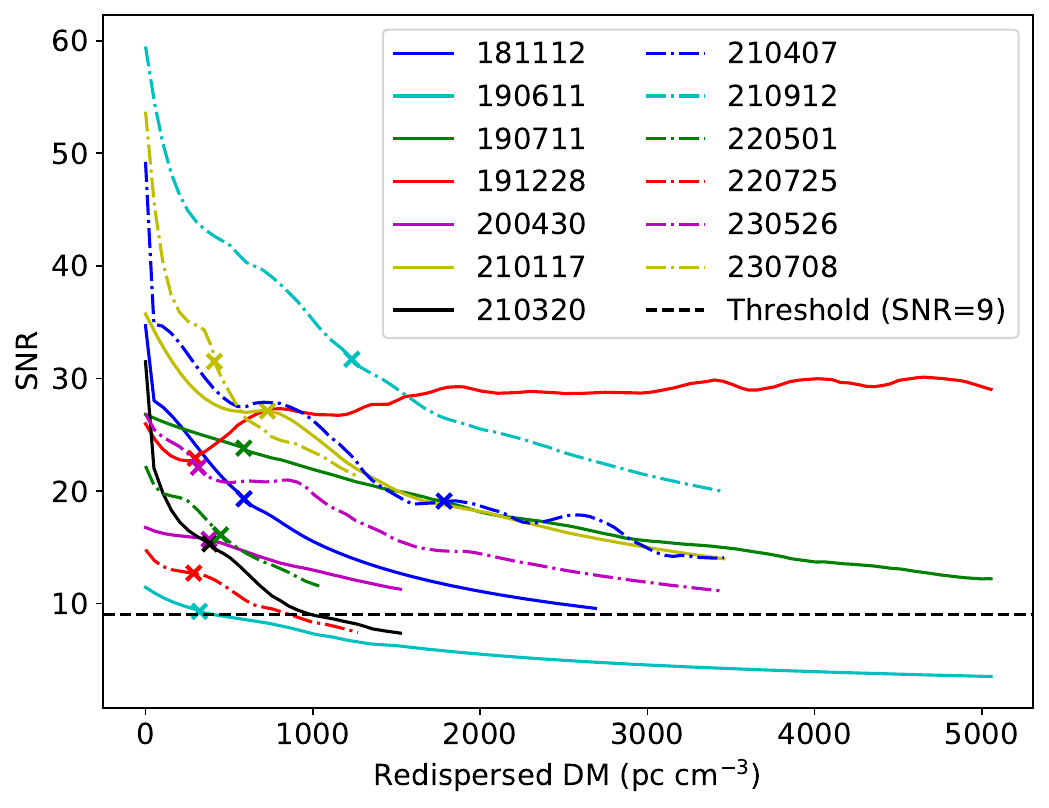}
\caption[]{The response of FREDDA at the time of detection for each FRB. Each curve is normalised to the corresponding cross which shows the DM and SNR at which the FRB was detected. The horizontal black dashed line shows the SNR threshold of FREDDA. The largest DM for each curve corresponds to the 4096$^{\rm th}$ time sample for the given FRB parameters which is the maximum searched DM of FREDDA. These curves are scaled to give $\eta$(DM) for use in predicting $H_0$.}
\label{fig:all_smoothed}
\end{center}
\vspace{-3ex}
\end{figure}

\citet{james2022b} measured $H_0$ by fitting cosmological and FRB population parameters to observed FRB characteristics. Their work models $\eta$(DM) using the \citet{Cordes2003} approximation. This does not consider an FRB-specific response and hence introduces a systematic error. We aim to quantify the impact that such an approximation has on estimations of $H_0$ with the sample of FRBs processed by CELEBI at the time of publication. To do so, we repeat the analysis of \citet{james2022b} with the 13 localised FRBs presented in this paper while only allowing $H_0$ to vary. The relevant observational parameters for such an analysis are given in Table \ref{tab:FRB_properties}. \myedit{We use the same cosmological assumptions and models presented in \citet{james2022} and \citet{james2022b} which are based upon a flat $\Lambda$CDM cosmology and use cosmological parameters measured by Planck \citep[][]{Planck2018}. In particular, we note that this tightly constrains the value of $\Omega_{\rm b} H_0^2$, which is what allows measurements of FRB DMs and redshifts (which constrain $\Omega_{\rm b} H_0$) to infer the value of $H_0$.} We then complete the same calculations while using the specific $\eta$(DM) curves for each FRB which we derive in Section \ref{sec:models}. To determine the response functions most accurately, we use the version of FREDDA which was operational at the time of detection. For CRAFT observations, FREDDA is set to search dispersions of up to 4096 time samples and thus we use this as the maximum DM value. This method additionally accounts for the finite search space of the detection algorithm which was not considered in previous analyses. The response functions are shown in Figure \ref{fig:all_smoothed}, normalised to the SNR of detection at the DM of detection. Normalisation is necessary as the HTR data is coherently beamformed while the real-time detection system operates on incoherently beamformed data and hence we expect an increase in the SNR proportional to $\sqrt{N_{\mathrm{antennas}}}$. In most instances, the gain is not as large as expected due to imperfect coherence. CRAFT observations have used a SNR threshold of 9 which is shown as the black dashed line. 

\begin{figure}
\begin{center}
\includegraphics[width=\columnwidth]{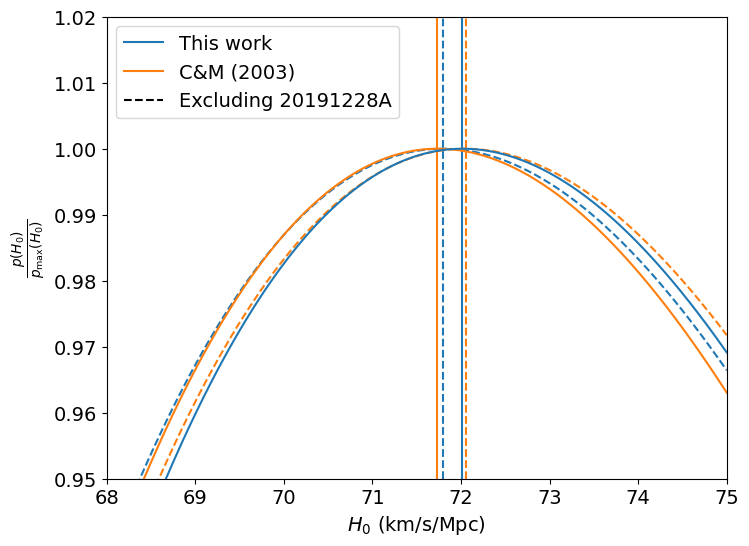}
\caption[]{The normalised posterior distribution on $H_0$ when the analysis is completed using (1) the specific FRB-by-FRB response (orange) and (2) using the original method of \citet{james2022b} which implements the \citet{Cordes2003} model (blue). The dashed lines show the same results excluding FRB 20191228A, which was detected with `Version 2' of FREDDA. The vertical lines show the maxima of each distribution located at 71.7, 71.8, 72.0 and 72.1$\,$km$\,$s$^{-1}\,$Mpc$^{-1}$.}
\label{fig:H0}
\end{center}
\vspace{-3ex}
\end{figure}

Figure \ref{fig:H0} shows the posterior distributions on $H_0$ thus obtained. When all of the FRBs in the dataset are considered, the calculation using the original method with the \citet{Cordes2003} model gives a best-fit value of 71.7$\,$km$\,$s$^{-1}\,$Mpc$^{-1}$ while a calculation using a specific FRB-by-FRB response gives a value of 72.0$\,$km$\,$s$^{-1}\,$Mpc$^{-1}$. FRB 20191228A has a significant individual contribution to the reported value of $H_0$. This FRB was detected with `Version 2' of FREDDA and hence the actual sensitivity function of the FRB is significantly different from the \citet{Cordes2003} model otherwise used. The real sensitivity function is artificially more sensitive to high DM values and therefore effectively decreases the probability of detecting low DM FRBs. As we do not detect as many high DM FRBs as would otherwise be suggested by such a sensitivity function, we prefer a universe that will produce lower DM values for a given redshift. Such a universe will be less dense. For this analysis, we keep $\Omega_{\rm b} H_0^2$ constant and thus this corresponds to a larger value of $H_0$. Thus, when excluding this FRB from the analysis, the best-fit $H_0$ reduces from 72.0$\,$km$\,$s$^{-1}\,$Mpc$^{-1}$ to 71.8$\,$km$\,$s$^{-1}\,$Mpc$^{-1}$ using the FRB-by-FRB response, but increases from 71.7$\,$km$\,$s$^{-1}\,$Mpc$^{-1}$ to 72.1$\,$km$\,$s$^{-1}\,$Mpc$^{-1}$ using the original method. Therefore, this individual FRB shifts the preferred value of $H_0$ by 0.6$\,$km$\,$s$^{-1}\,$Mpc$^{-1}$. This version also introduced difficulties in triggering detections due to a large number of high-DM candidates which is a more complex systematic to quantify. Thus, we exclude surveys completed with `Version 2' of FREDDA in future analyses. With such an exclusion, the FRB-specific response is 0.3$\,$km$\,$s$^{-1}\,$Mpc$^{-1}$ less than the results using the \citet{Cordes2003} model.

The source of this 0.3$\,$km$\,$s$^{-1}\,$Mpc$^{-1}$ difference is not only associated with the FRB-specific response. When replicating the analysis of \citet{james2022b}, we use an assumed $w_{\mathrm{inc}}$ value obtained from Gaussian fits of the bursts while the analysis using the FRB-specific responses uses optimised widths. If we complete the original analysis using the optimised widths as well then the systematic difference decreases to 0.1$\,$km$\,$s$^{-1}\,$Mpc$^{-1}$. Additionally, the original analysis does not consider the search limits of FREDDA in DM-space while the new analysis does. However, incorporating these search limits has minimal impact on the estimated value of $H_0$.

The difference of 0.3$\,$km$\,$s$^{-1}\,$Mpc$^{-1}$ is significantly less than the current statistical uncertainties from \citet{james2022b} which are on the order of $\sim$10$\,$km$\,$s$^{-1}\,$Mpc$^{-1}$. \citet{Planck2018} report a measured $H_0$ value of $67.4 \pm 0.5\,$km$\,$s$^{-1}\,$Mpc$^{-1}$ and \citet{SH0ES2021} report a value of $73.04 \pm 1.04$$\,$km$\,$s$^{-1}\,$Mpc$^{-1}$. Systematic errors on the order of $\sim$0.3$\,$km$\,$s$^{-1}\,$Mpc$^{-1}$ will therefore be significant when the statistical uncertainty is reduced to a comparable level. \citet{james2022b} estimate statistical uncertainties for 100 localised FRBs to be 2.45$\,$km$\,$s$^{-1}\,$Mpc$^{-1}$ and 1.2$\,$km$\,$s$^{-1}\,$Mpc$^{-1}$ for 400 localised FRBs. Hence, for this sample size, systematics due to approximations of $\eta$(DM) should be considered.


\subsection{Varying additional parameters}
\myedit{Fits of $H_0$ also require simultaneously fitting FRB population parameters, with which $H_0$ is correlated. In \citet{james2022b}, they consider the parameters $n_{\rm{sfr}}$ which is defined such that the rate of FRBs, $\Phi$, scales with the star formation rate of the Universe, SFR($z$), as $\Phi(z) \propto \rm{SFR}(z)^{n_{\rm{sfr}}}$; the spectral index, $\alpha$, defined such that $\Phi(\nu) \propto \nu^{\alpha}$; the mean, $\mu_{\rm{host}}$, and standard deviation, $\sigma_{\rm{host}}$, of the DM contributions from host galaxies; the maximum energy of an FRB, $E_{\rm{max}}$; and the slope of the integrated luminosity function, $\gamma$. Thus, a more accurate reflection of the systematic effects on $H_0$ would include these correlations in a multi-parameter fit. However, with the current implementation, it is not computationally feasible to allow all of the parameters to vary using the FRB-specific responses. We have therefore implemented an MCMC sampler to explore the parameter space more efficiently. The full details of this will be presented in a future analysis (\hoffmanninprep). This implementation has allowed us to vary all of the relevant parameters of the model -- that is, the parameters which are shown in Table \ref{tab:systematics} which correspond to the free parameters of \citet{james2022b}. However, MCMC sampling has intrinsic variability in the resulting distributions. This variability is difficult to disentangle from the systematic which we aim to investigate and hence we do not use the results of this section as the main result of the paper.}

\myedit{Due to the small sample size that we have available for this analysis, many of the parameters are weakly constrained. As such, we do not place great emphasis on the absolute values of each of the parameters. A more detailed analysis with a larger sample size will be presented at a later date. Here, we present the systematic differences in each of the parameters when introducing the FRB-specific response curves. When comparing numerical values, we characterise the difference in a parameter using $\Delta_{\rm{val}} = x_{\rm{exact}} - x_{\rm{C\&M}}$, where $x$ is the median of the posterior distributions, `exact' refers to the estimation using the FRB-specific response curves from this work and `C\&M' refers to the analysis with the \citet{Cordes2002} model. We do this because the median is the most stable quantifier.}

\myedit{In an effort to isolate the systematic of interest, we attempt to quantify fluctuations in the determined median value which are caused by intrinsic variations from the MCMC method. The analysis is computationally very expensive and hence it is impractical to repeat the full analysis multiple times to obtain an accurate empirical result. Instead, we take equally sized subsets of the MCMC sample and determine the median values in each subset. For each parameter, we then determine the standard deviation for the set of median values. By repeating this for varying subset sizes we can obtain a trend that shows a decrease in the standard deviation proportional to the square root of the subset size. This is extrapolated to the total sample size to give an estimate of the variation due to the MCMC sampler $\sigma_{\rm{MCMC}}$. These values are approximate and not rigorous, but should nevertheless give a characteristic estimate for these uncertainties.}

\begin{table}
\begin{center}
\caption{\myedit{Systematic differences $\Delta_{\rm{val}}$ when allowing all parameters to vary. $\sigma_{\mathrm{MCMC}}$ is an approximation of the 1-$\sigma$ random uncertainties from MCMC sampling. The parameters presented here are the free parameters which are allowed to vary according to the analysis of \citet{james2022b}.}}
\label{tab:systematics}
\begin{tabular}{lll}
\hline
Parameter \hspace{1ex} & $\Delta_{\mathrm{val}}$ \hspace{1ex} & $\sigma_{\mathrm{MCMC}}$ \\
\hline 
\vspace{-2ex} \\
$n_{\mathrm{sfr}}$ & -0.137 & 0.025 \\
$\alpha$ & -0.829 & 0.014 \\ 
log$_{10}$($\mu_{\rm{host}}$) & -0.077 & 0.007 \\ 
log$_{10}$($\sigma_{\rm{host}}$) & 0.048 & 0.016 \\ 
log$_{10}$(E$_{\mathrm{max}}$) & 0.097 & 0.008 \\ 
$\gamma$ & -0.057 & 0.012 \\ 
H$_0$ & -1.32 & 0.23 \\
\hline
\end{tabular}
\end{center}
\end{table}

\myedit{Table \ref{tab:systematics} shows a summary of the systematic differences introduced when considering FRB-specific responses. The value of $\alpha$ shows the most significant changes.}

\myedit{When using a multi-parameter fit, the use of FRB-specific response functions causes a decrease in $H_0$ of $\sim$1$\,$km$\,$s$^{-1}\,$Mpc$^{-1}$ as opposed to the noted decrease of $\sim$0.3$\,$km$\,$s$^{-1}\,$Mpc$^{-1}$ when only allowing $H_0$ to vary. This does indicate that the systematics on $H_0$ could be larger than initially estimated and hence would become more significant in resolving the Hubble Tension. In both instances, the inclusion of FRB-specific response functions causes a decrease in the value of $H_0$ therefore suggesting that previous analyses were biased towards higher values.}


\section{Conclusions} \label{sec:conclusions}
In this work, we coherently redisperse HTR data of 13 CRAFT-detected FRBs to empirically determine the sensitivity of FREDDA -- the CRAFT FRB detection system. We compare our results with theoretical predictions presented in \citet{Cordes2003} and \citet{Arcus2020} as well as the injection of idealised pulses as in \citet{Qiu2023}. 

We find that provided the incident width of the FRB $w_{\mathrm{inc}}$ is optimised, the results of the idealised pulse injection describe the sensitivity function well. As such, we conclude that for most FRBs, the temporal and spectral structures have minimal impacts on the sensitivity function at typical temporal and spectral resolutions of FREDDA. The exceptions to this are FRB 20190711A, which is broad in time and only fills the lower half of the band, and FRB 20230708A, which shows a transition in the incident width from 1.5 ms to 10.0 ms. We also characterise the response of two development versions of FREDDA, finding that `Version 2' used for FRB 20191228A produced artificially high SNR values at high DM.

We additionally investigate the impacts of approximating the sensitivity function using a theoretical model in calculations of $H_0$. We find that FRB 20191228A creates a systematic error of 0.6$\,$km$\,$s$^{-1}\,$Mpc$^{-1}$ due to it being detected with `Version 2' of FREDDA. As such, excluding surveys completed with `Version 2' will help minimise systematic errors. With the exclusion of this FRB, using our empirical FRB-by-FRB response for the sample of FRBs currently processed by CELEBI creates a difference of 0.3$\,$km$\,$s$^{-1}\,$Mpc$^{-1}$ \myedit{when only allowing $H_0$ to vary}. This difference is partially (0.1$\,$km$\,$s$^{-1}\,$Mpc$^{-1}$) attributed to using optimised widths; accounting for the search limits of FREDDA shows little impact. \myedit{We also note that this systematic may be larger on the order of $\sim1\,$km$\,$s$^{-1}\,$Mpc$^{-1}$ if all other parameters are allowed to vary.} Overall, this systematic is currently irrelevant due to the statistical uncertainties on our estimations of $H_0$. However, we hope to be able to resolve the Hubble tension with $\sim\,$400 localised FRBs at which point this effect will be significant.

\section{Acknowledgements}
We thank the referee for their input and insightful comments. We also thank E. Keane for his input on the manuscript.

This work was performed on the OzSTAR national facility at Swinburne University of Technology. The OzSTAR programme receives funding in part from the Astronomy National Collaborative Research Infrastructure Strategy (NCRIS) allocation provided by the Australian Government. This scientific work uses data obtained from Inyarrimanha Ilgari Bundara, the CSIRO Murchison Radio-astronomy Observatory. We acknowledge the Wajarri Yamaji as the Traditional Owners and native title holders of the Observatory site. CSIRO’s ASKAP radio telescope is part of the Australia Telescope National Facility. The operation of ASKAP is funded by the Australian Government with support from the National Collaborative Research Infrastructure Strategy. ASKAP uses the resources of the Pawsey Supercomputing Research Centre. The establishment of ASKAP, Inyarrimanha Ilgari Bundara, the CSIRO Murchison Radio-astronomy Observatory and the Pawsey Supercomputing Research Centre are initiatives of the Australian Government, with support from the Government of Western Australia and the Science and Industry Endowment Fund.

CWJ and MG acknowledge support through Australian Research Council (ARC) Discovery Project (DP) DP210102103. ATD acknowledges support through ARC DP DP220102305. KG acknowledges support through ARC DP DP200102243. RMS acknowledges support through Australian Research Council Future Fellowship F190100155 and Discovery Project DP220102305.
 J.X.P. and N.T. acknowledge support from NSF grants AST-1911140, AST-1910471
and AST-2206490 as members of the Fast and Fortunate for FRB
Follow-up team.

\section{Data Availability}
A repository of the code used in this study can be found at \href{https://github.com/JordanHoffmann3/RedisperseFRB}{github.com/JordanHoffmann3/RedisperseFRB}. This analysis also uses code from \citet{james2022b} which is available on GitHub \citep{frb,zdm}. The data used is multiple GB but can be made available upon request to the authors.

\bibliography{References}
\bibliographystyle{mnras}

\appendix
\section{Sensitivity curves}
\begin{figure*}
\Large{FRB 20181112A}
\begin{subfigure}{8.7cm}
  \centering
  \includegraphics[width=8.7cm]{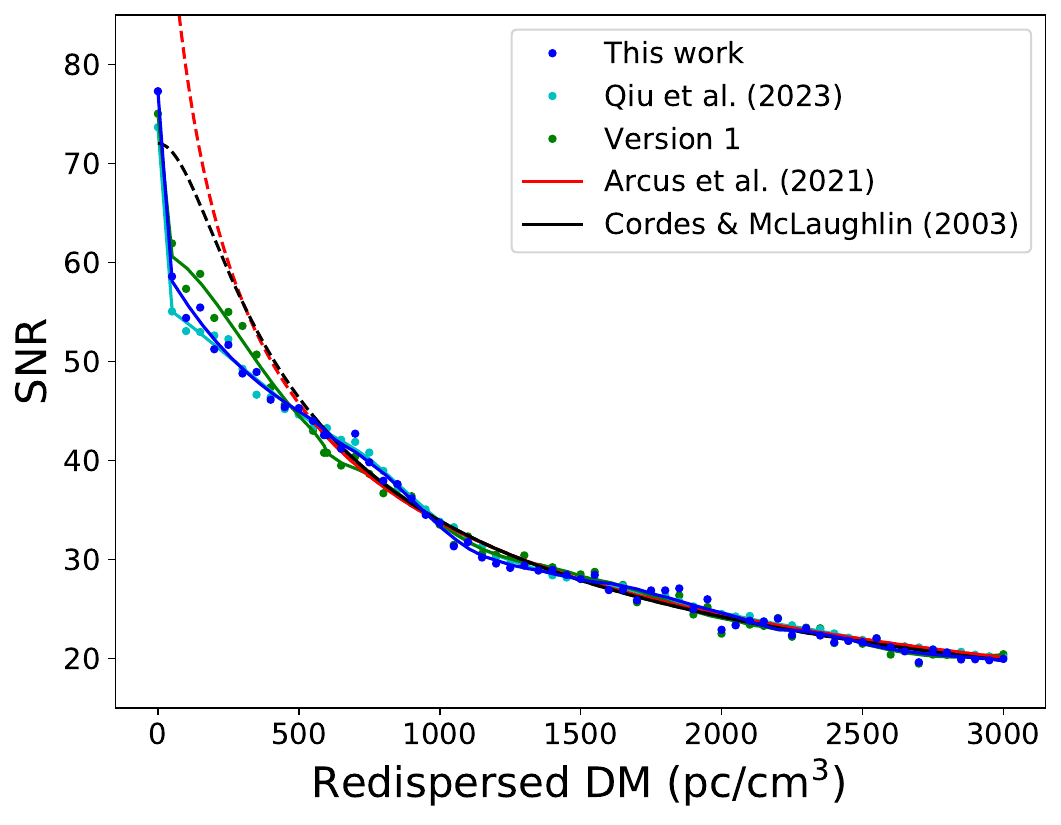}
\end{subfigure}%
\begin{subfigure}{7.3cm}
  \centering
  \includegraphics[width=7.3cm]{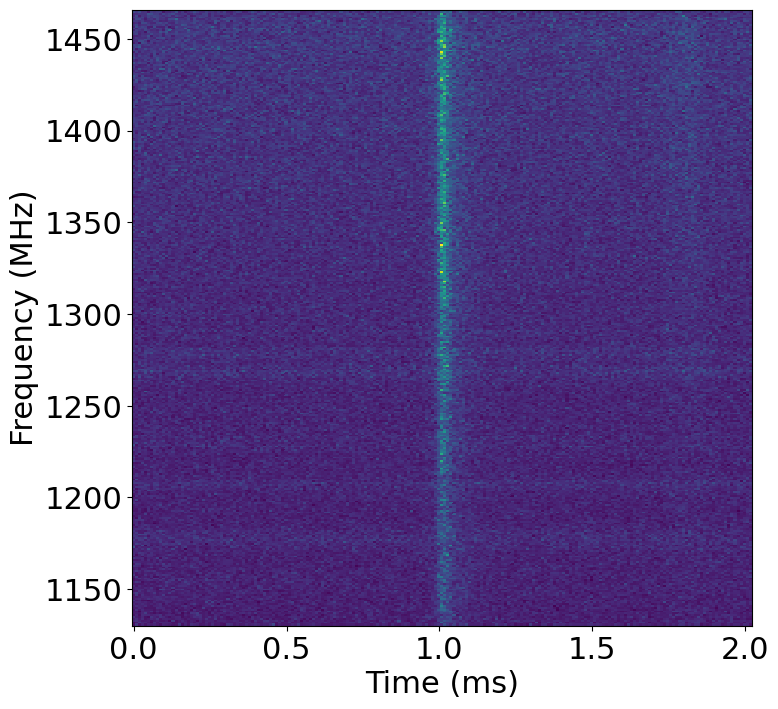}
\end{subfigure}

\Large{FRB 20190611B}
\begin{subfigure}{8.7cm}
  \centering
  \includegraphics[width=8.7cm]{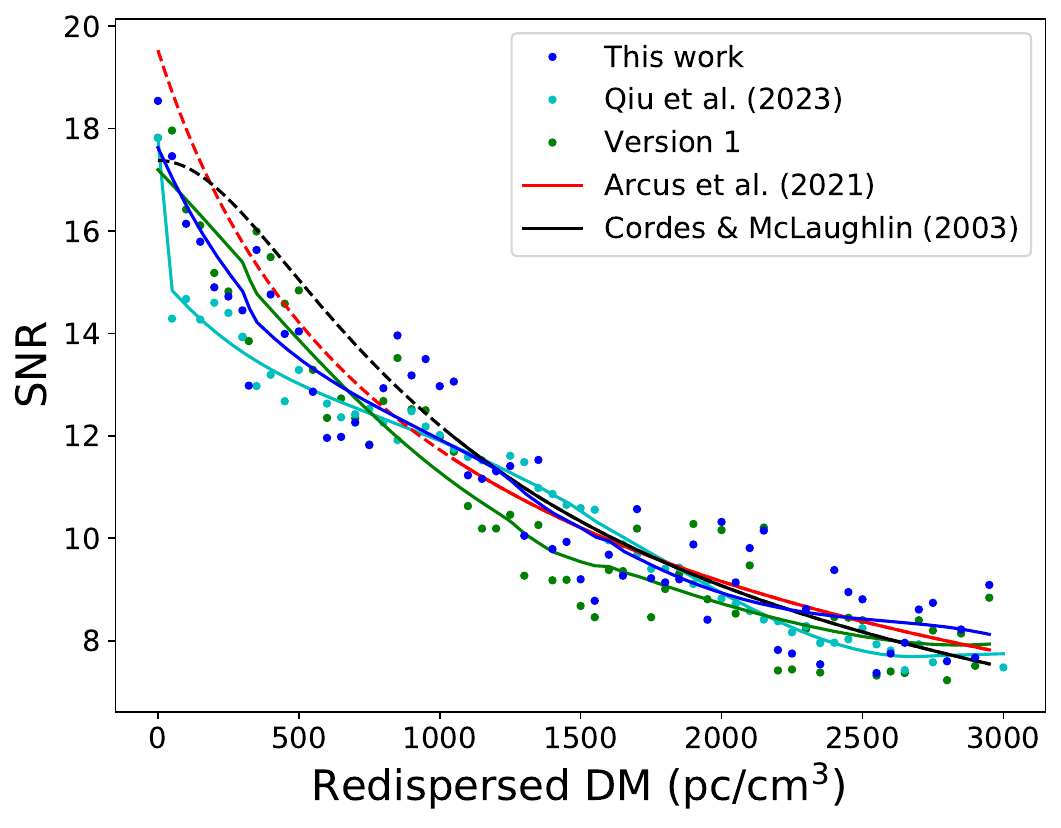}
\end{subfigure}%
\begin{subfigure}{7.3cm}
  \centering
  \includegraphics[width=7.3cm]{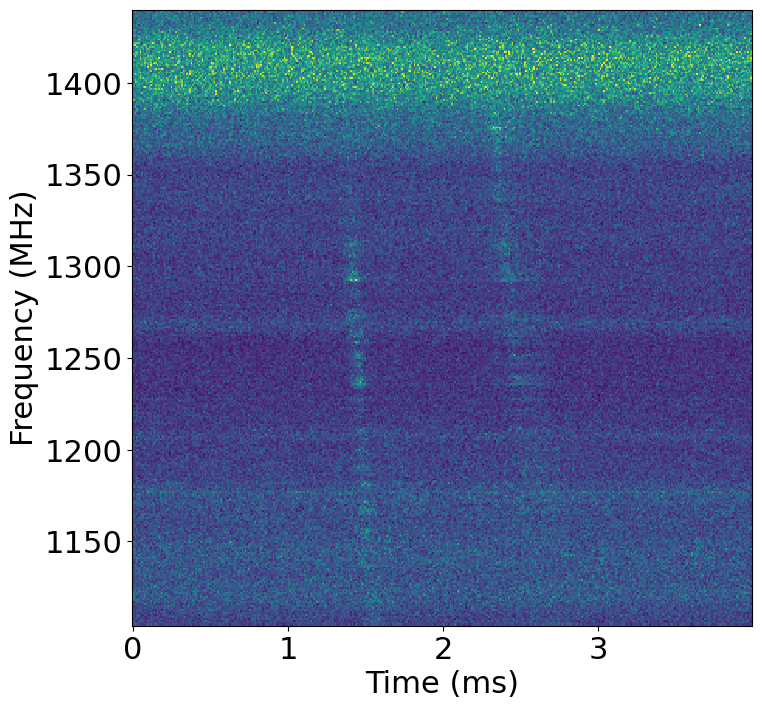}
\end{subfigure}

\caption{Catalogue of the sensitivity functions for each of the FRBs on which the analysis was completed. Each plot replicates Figure \ref{fig:220501_SNR} for a different FRB and hence a more detailed explanation of the represented data is given in Section \ref{sec:models}. The plots shown in the main body are not repeated here. In brief, we show our results (dark blue), idealised pulse injection of \citet{Qiu2023} (light blue), results using the version of FREDDA running at the time of detection (green), the best-fit model of \citet{Arcus2020} (red) and the best-fit model of \citet{Cordes2003} (black).}
\label{fig:SNRs}
\end{figure*}

\begin{figure*} \ContinuedFloat
\Large{FRB 20191228A}
\begin{subfigure}{8.7cm}
  \centering
  \includegraphics[width=8.7cm]{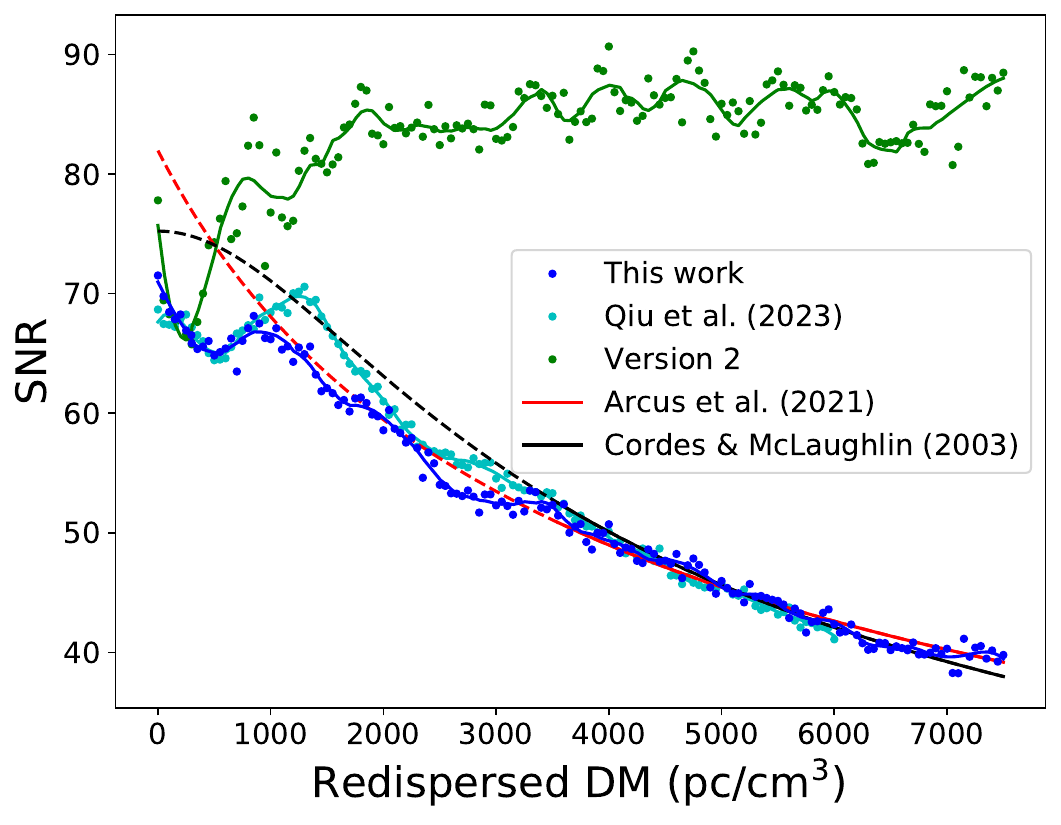}
\end{subfigure}%
\begin{subfigure}{7.3cm}
  \centering
  \includegraphics[width=7.3cm]{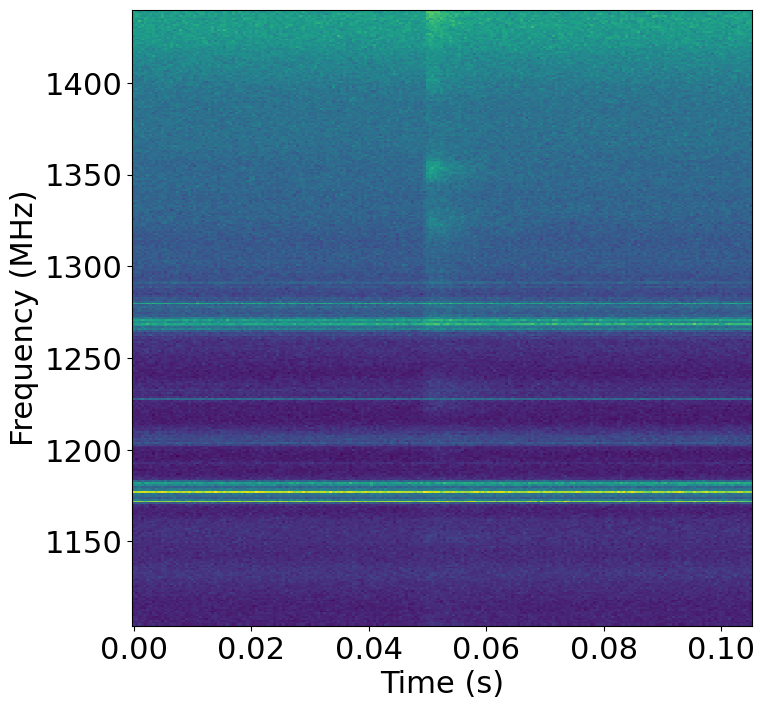}
\end{subfigure}

\Large{FRB 20200430A}
\begin{subfigure}{8.7cm}
  \centering
  \includegraphics[width=8.7cm]{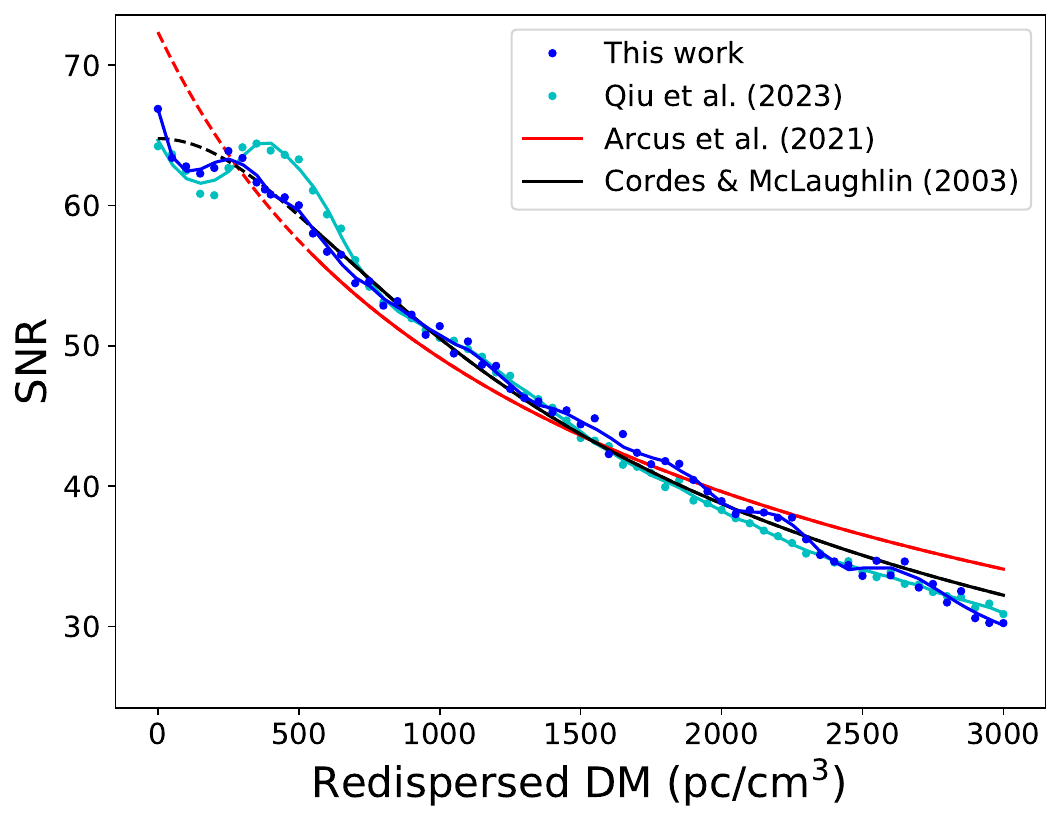}
\end{subfigure}%
\begin{subfigure}{7.3cm}
  \centering
  \includegraphics[width=7.3cm]{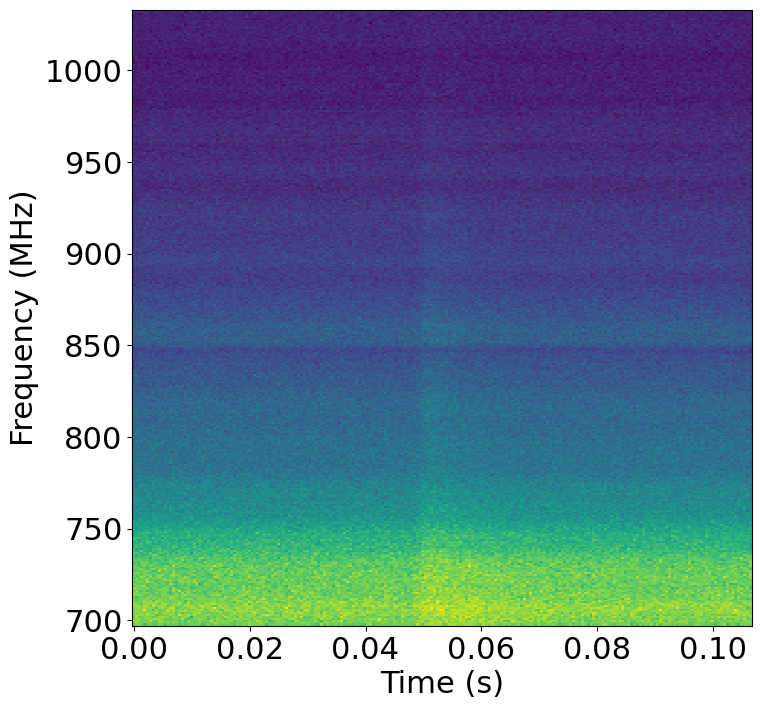}
\end{subfigure}

\Large{FRB 20210117A}
\begin{subfigure}{8.7cm}
  \centering
  \includegraphics[width=8.7cm]{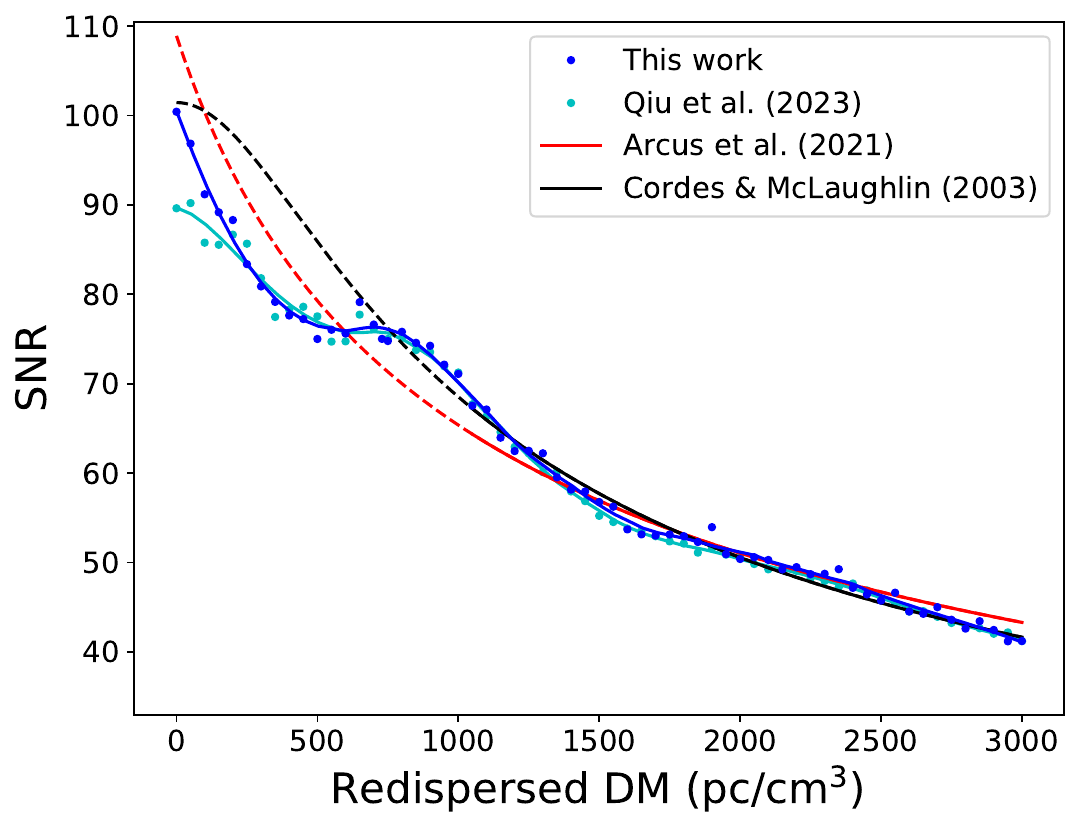}
\end{subfigure}%
\begin{subfigure}{7.3cm}
  \centering
  \includegraphics[width=7.3cm]{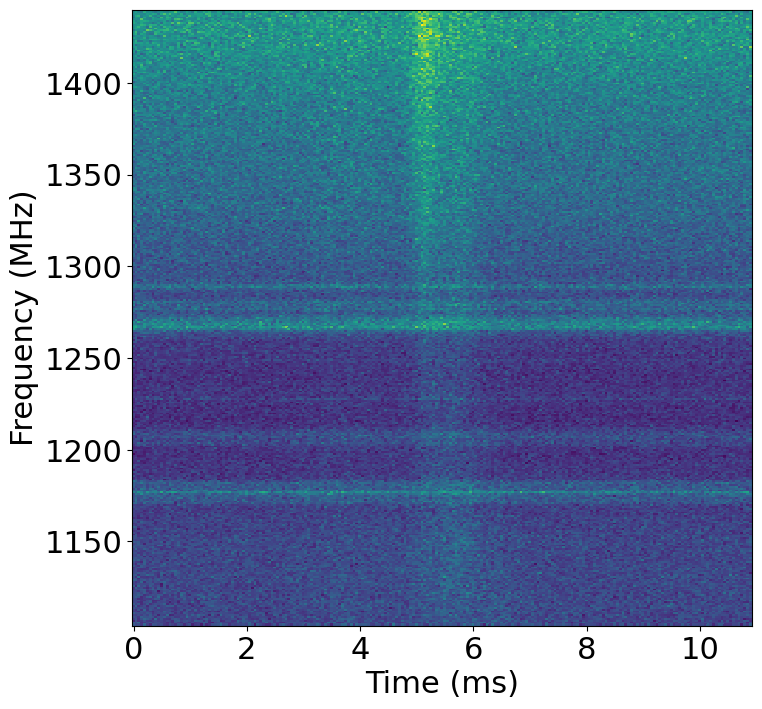}
\end{subfigure}

\caption{--- \textit{continued}. Where no green points are present the latest version of FREDDA was operational at detection which was used to produce the dark blue results.}
\end{figure*}

\begin{figure*} \ContinuedFloat
\Large{FRB 20210320A}
\begin{subfigure}{8.7cm}
  \centering
  \includegraphics[width=8.7cm]{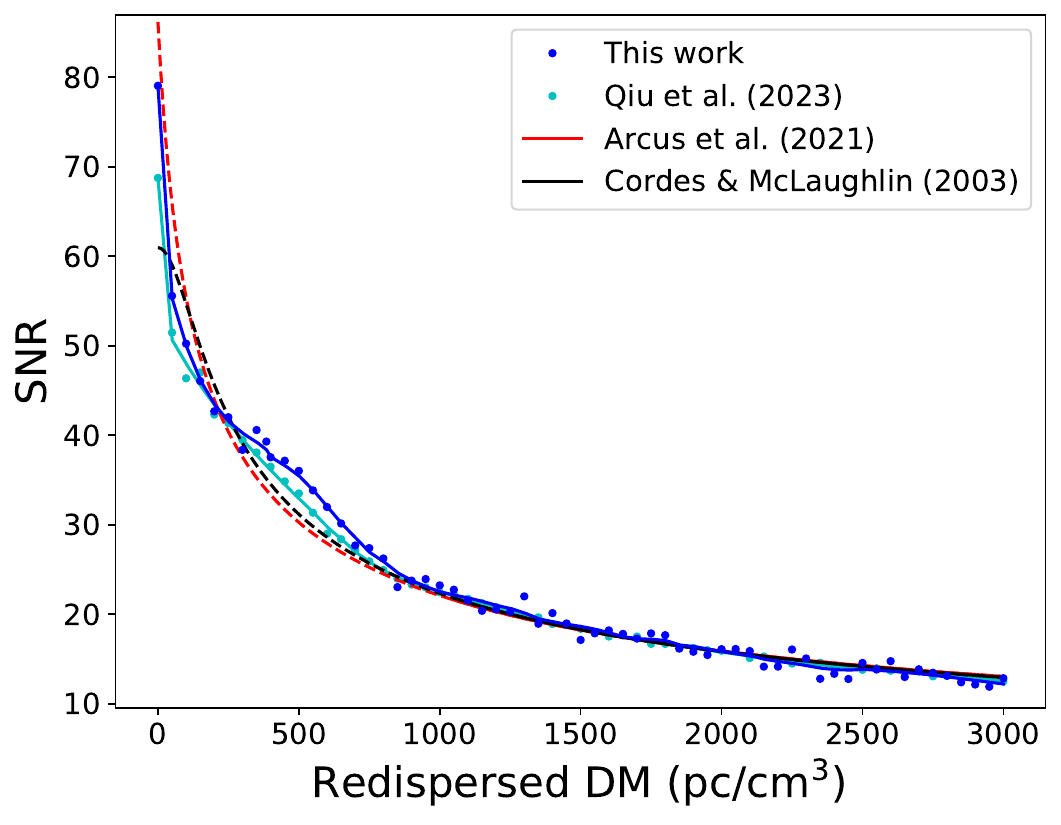}
\end{subfigure}%
\begin{subfigure}{7.3cm}
  \centering
  \includegraphics[width=7.3cm]{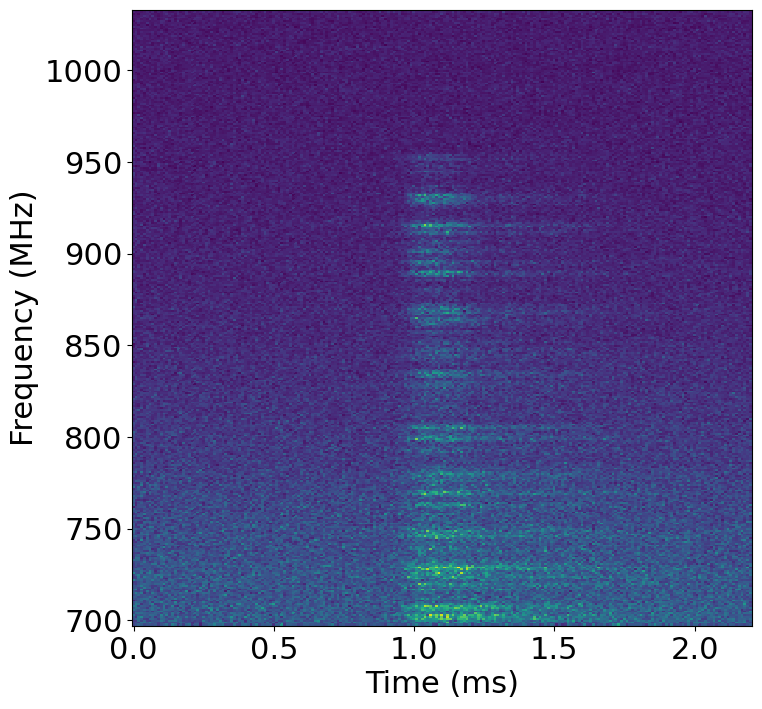}
\end{subfigure}

\Large{FRB 20210407E}
\begin{subfigure}{8.7cm}
  \centering
  \includegraphics[width=8.7cm]{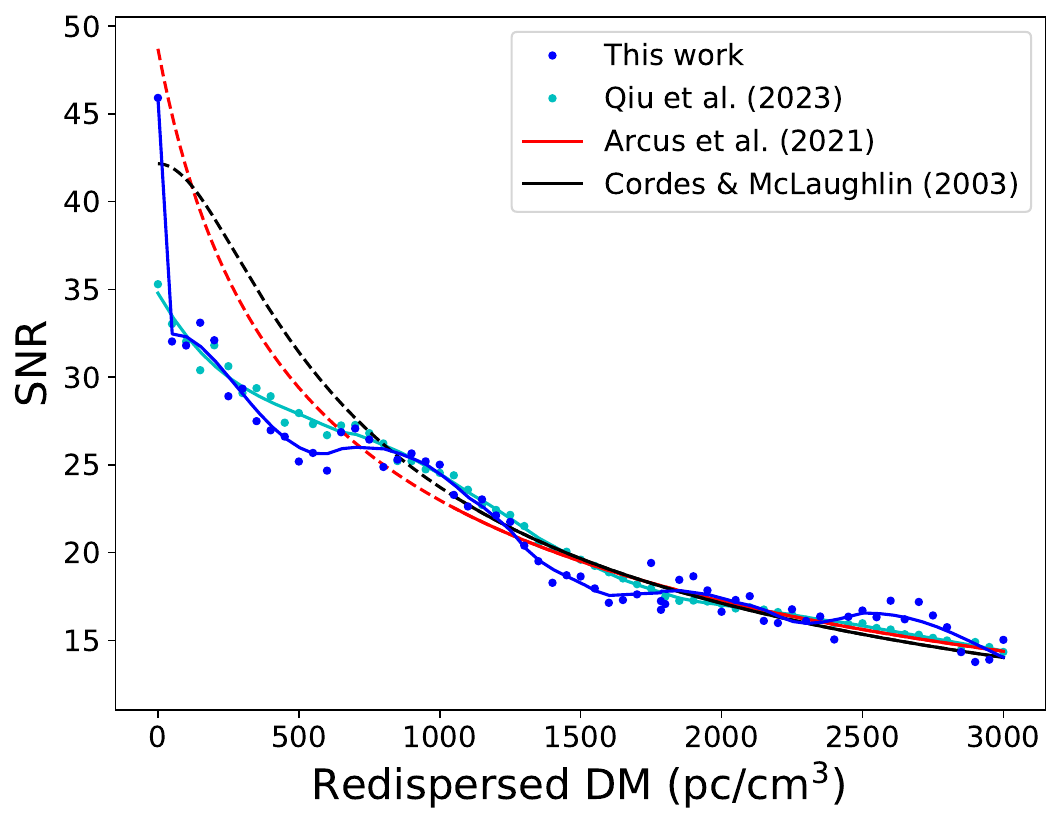}
\end{subfigure}%
\begin{subfigure}{7.3cm}
  \centering
  \includegraphics[width=7.3cm]{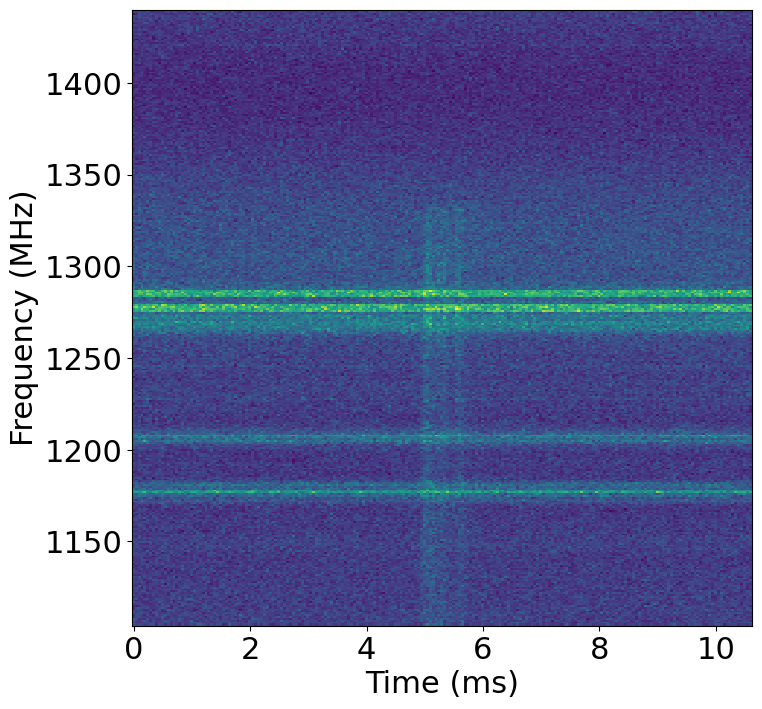}
\end{subfigure}

\Large{FRB 20210912A}
\begin{subfigure}{8.7cm}
  \centering
  \includegraphics[width=8.7cm]{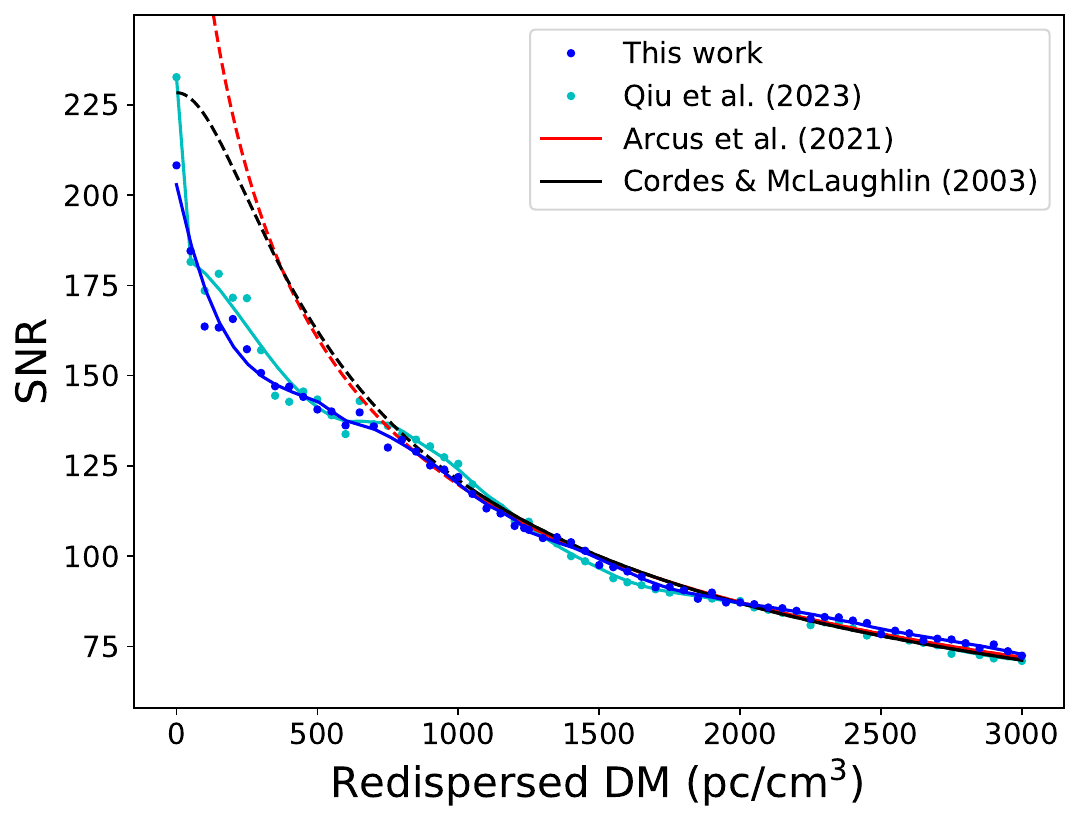}
\end{subfigure}%
\begin{subfigure}{7.3cm}
  \centering
  \includegraphics[width=7.3cm]{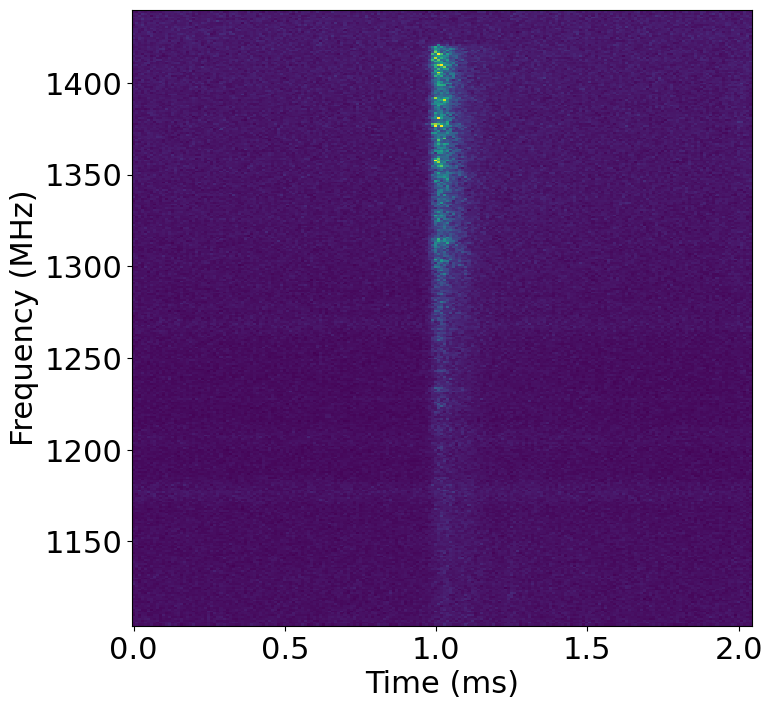}
\end{subfigure}

\caption{--- \textit{continued}.}
\end{figure*}

\begin{figure*} \ContinuedFloat
\Large{FRB 20220725A}
\begin{subfigure}{8.7cm}
  \centering
  \includegraphics[width=8.7cm]{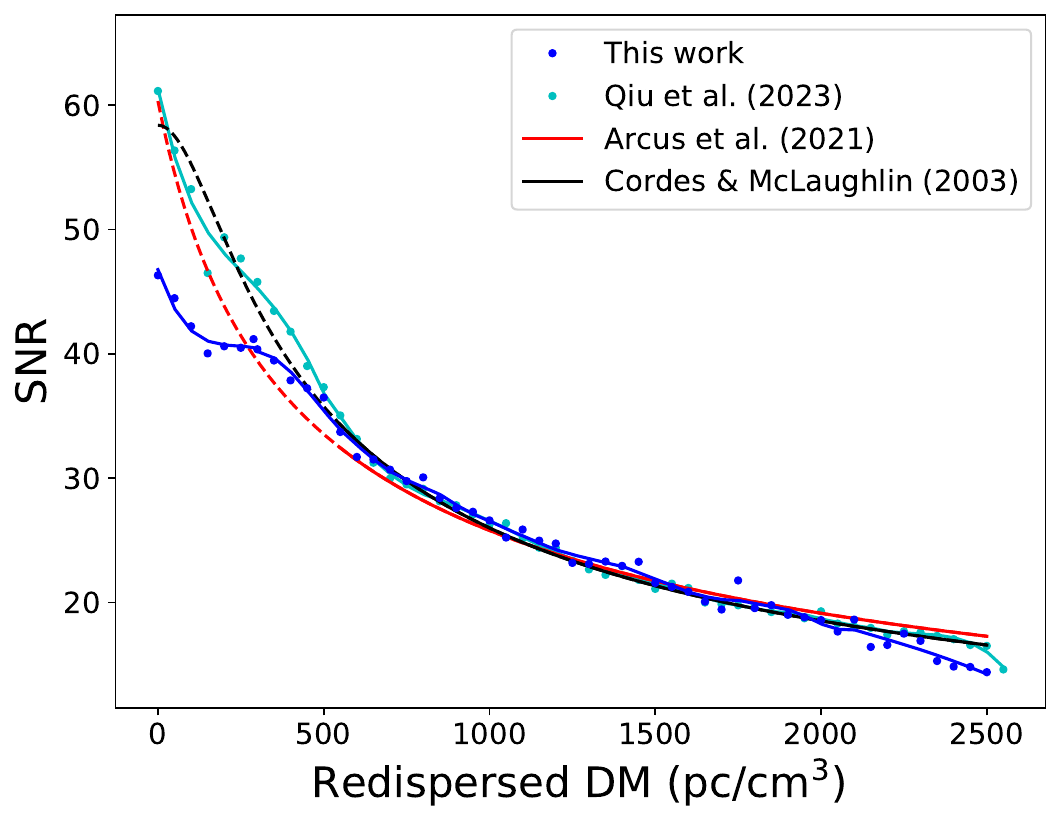}
\end{subfigure}%
\begin{subfigure}{7.3cm}
  \centering
  \includegraphics[width=7.3cm]{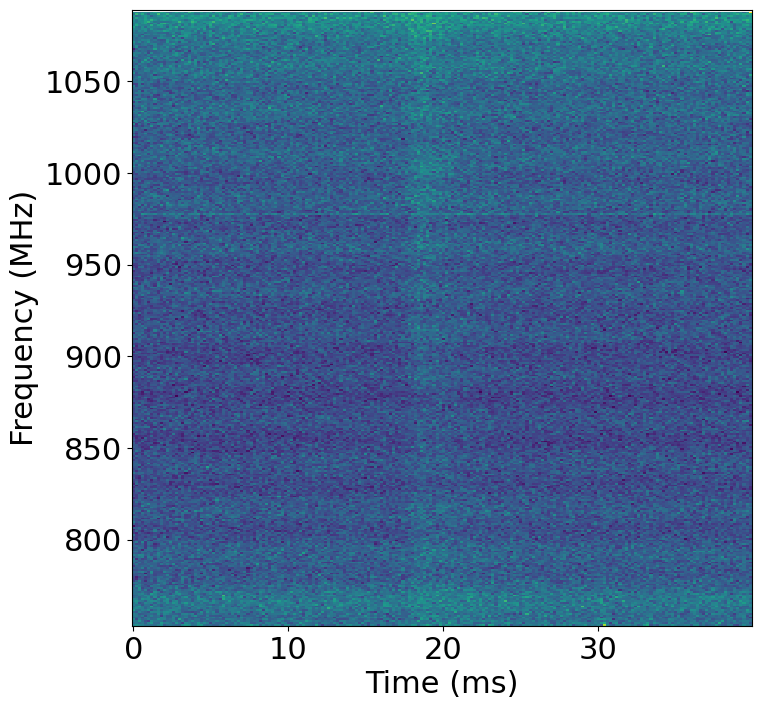}
\end{subfigure}

\Large{FRB 20230526A}
\begin{subfigure}{8.7cm}
  \centering
  \includegraphics[width=8.7cm]{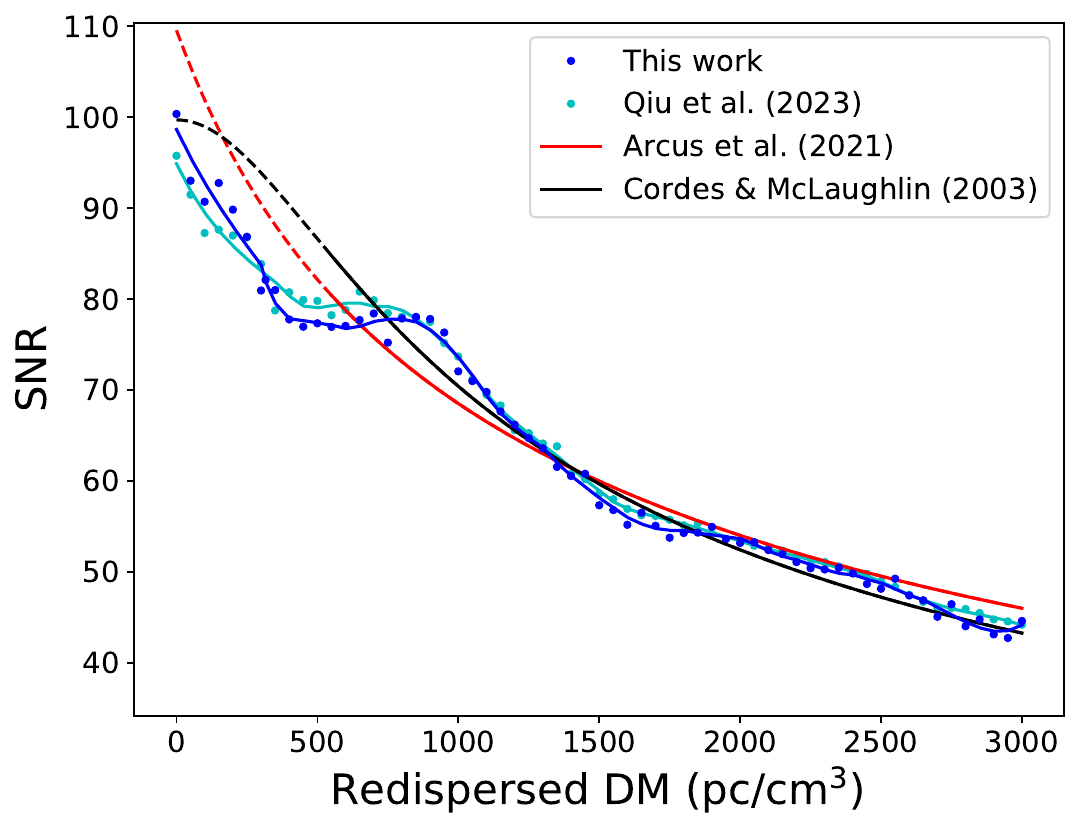}
\end{subfigure}%
\begin{subfigure}{7.3cm}
  \centering
  \includegraphics[width=7.3cm]{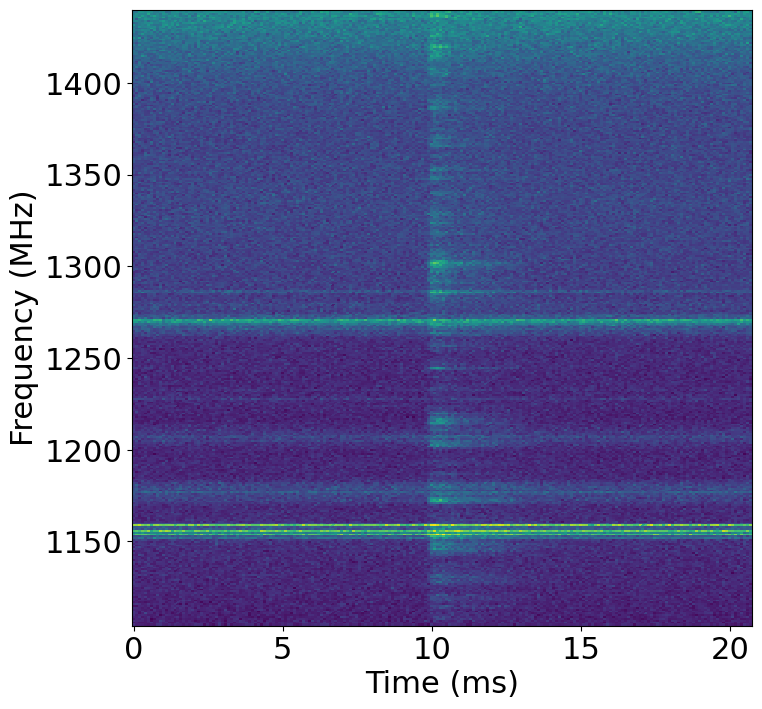}
\end{subfigure}

\caption{--- \textit{continued}. The sinusoidal modulation in the dynamic spectrum of FRB 20220725A can be attributed to reflections between the receiver and dish which are $\sim$6 m apart and hence a $\sim$12 m pathlength corresponds to constructive interference for frequencies that are a multiple of $\sim$25 MHz.}
\end{figure*}

\end{document}